\documentclass{aa}
\usepackage{txfonts}
\usepackage{graphicx}
\usepackage[round]{natbib}

\def\Sch{Schr{\"o}dinger }
\def\d{{\rm d}}

\begin{document}

   \title{Cosmological neutrino entanglement and quantum pressure}

    \author{D.~Pfenniger \and V.~Muccione}

    \institute{Geneva Observatory, University of Geneva, 1290
               Sauverny, Switzerland }

    \date{Received 09/11/2005 / Accepted 03/06/2006}

\abstract
%Context
{ The widespread view that cosmological neutrinos, even if massive,
  are well described since the decoupling redshift $z \approx 10^{10}$
  down to the present epoch by an almost perfectly
  \textit{collisionless} fluid of classical point particles is
  re-examined.  }
%Aims
{ In view of the likely sub-eV rest mass of neutrinos, the main 
  effects due to their fermionic nature are studied.  }
%Methods
{ By numerical means we calculate the accurate entropy, fugacity and
  pressure of cosmological neutrinos in the Universe expansion.  By
  solving the \Sch equation we derive how and how fast semi-degenerate
  identical free fermions become entangled.  }
%Results
{ We find that for sub-eV neutrinos the exchange degeneracy has
  significantly \textit{increased} during the relativistic to
  non-relativistic transition epoch at $z\approx 10^4-10^5$.  At all
  times neutrinos become entangled in less than $10^{-6}\,$s, much
  faster than any plausible decoherence time.  The total pressure is
  increased by quantum effect from 5\% at high redshifts to 68\% at
  low redshifts with respect to a collisionless classical fluid.  }
%Conclusions
{ The quantum overpressure has no dynamical consequences in the
  homogeneous regime at high redshifts, but must be significant for
  neutrino clustering during the non-linear structure formation epoch
  at low redshifts.  } 


\maketitle

%________________________________________________________________

\section{Introduction}

We re-examine the common view in astronomy that cosmological
neutrinos, if massive, are well described since their decoupling
redshift $z \approx 10^{10}$ down to the present epoch by a
collisionless fluid of classical particles \citep{primack00}.
Neglected up to now is one of the most fundamental and well
established principle of modern physics, the fermion/boson
symmetrisation rules from which Pauli's exclusion principle follows
\citep{pauli25, dirac26, fermi26}.

When we asked cosmologists how massive neutrinos behave today despite
their degeneracy,\footnote{In the context of cosmological neutrinos,
  the word ``degeneracy'' is used with different meanings.  Here it
  concerns the particle exchange degeneracy, not the mass or the
  chemical potential degeneracy.} 
they would argue that they behave as a collisionless fluid of
classical particles. In cosmological N-body simulations
\citep[e.g.,][]{klypin93} the only difference between neutrinos and
cold dark matter particles is the respective initial velocity
dispersion.

When we asked theoretical and experimental physicists familiar with
particle physics, quantum and statistical mechanics, they would find
obvious that degeneracy pressure between neutrinos must be taken into
account despite their negligible weak interaction cross section,
because degeneracy pressure results from Pauli's principle and depends
only on the exchange symmetry of identical particles. In the limit of
negligible interaction in the particle Hamiltonian, the fermionic or
bosonic behaviour remains, therefore the particle statistics is an
aspect unrelated to the strength of the particle-particle interaction.

In summary, in the cosmologist view neutrinos are well described with
a distribution of \textit{collisionless} classical particles following
the collisionless Boltzmann equation, while in the physicist view
neutrinos are quantum particles, i.e., non-local entities, and when
they are indistinguishable they behave collectively according to the
well established principles of quantum statistics. In cosmological
simulations neutrinos would be better represented by a
\textit{collisional} gas with the Fermi-Dirac equation of state.

This distinction has no consequence during the relativistic
homogeneous regime because the initial Fermi-Dirac statistics is
preserved by the Universe expansion \citep{weinberg72}. But this is
no longer true during the non-linear structure formation phase, where
the different matter components induce inhomogeneous perturbations
driving neutrinos out of thermal equilibrium. Then the degeneracy
pressure is different from the straight kinetic pressure of
collisionless particles, therefore the fate of massive neutrinos in
galaxy clusters or galaxies differs if neutrinos behave as classical
or quantum particles.

In the recent years this problem has taken importance for the
description of the non-linear phase of structure formation, because
the likely mass of neutrinos \citep{tegmark05} determined in different
experiments appear to sum up to at least the
\textit{visible} baryon mass fraction.  If neutrinos behave
collectively as a classical collisionless fluid, then their
interaction with the other matter components is limited to the weak
and gravitational interactions, while if neutrinos behave as a
coherent Fermi gas, then they can develop shocks, increase entropy,
and lead to non-linear evolution as hard to predict as the baryon
one during the structure formation phase.

The paradox raised by the cosmological neutrinos goes in fact at the
heart of still debated fundamental open problems in quantum mechanics,
such as the meaning of measurements and wave-function reductions, the
destruction or built-up of coherence between particles, etc.  It
concerns also the problem of irreversibility in quantum systems,
which, as in classical systems, are strictly time-reversible in theory,
but dissipative in practice.

%_______________________________________________________________________

\section{The classical description}

It has been usual in the past decades to assume neutrinos as either
massless relativistic particles, or as a dark matter candidate able to
account for a substantial part of the dark matter, which implies a
neutrino rest mass well above 10\,eV.  The classical reasoning 
\citep[e.g.,][]{tg79, peebles93} goes as follow.  Neutrinos are
generated above redshift $z \approx 10^{10}$ by elementary particle
creation and destruction processes.  The important point is that at
these densities ($\sim 10^{31}\,\rm cm^{-3}$) and temperature ($\sim
10^{10}\, \rm K$) the electro-weak force cross-sections are large
enough to make the reaction time-scales much shorter than the
Universe expansion time-scale 
\citep[see][]{weinberg72, peebles93, pad02}.  
This ensures thermal equilibrium
and well determined initial thermodynamical conditions.  At redshift
below about $z \approx 10^{10}$ the creation of electron-positron
pairs, which releases the last neutrinos and anti-neutrinos, drops because 
the thermal energy of particles falls below the
$511\,\rm keV$ electron-positron rest mass energy.  The neutrino mean
free-path for weak interaction diverges and neutrinos are since then
considered as propagating freely across the Universe, only subject to
gravity perturbations.

The neutrino weak interaction cross section is exceedingly small
\citep{peacock99}:
\begin{equation}
\sigma \sim 4\cdot 10^{-64}\, T^2\, {\rm cm^2}\ .
\label{equ:WS}
\end{equation}
Therefore the widespread view among cosmologists is to consider
neutrinos as collisionless classical particles, i.e., well localised
mass concentrations moving according to Newton's laws.  Consequently,
neutrinos are supposed to obey accurately the collisionless Boltzmann
equation.  The neutrinos follow an incompressible flow in phase space,
even during the possibly complex non-linear structure formation phase.
This leads to the well known phase space density constraint
\citep{tg79}. In this classical description violent relaxation
processes dilute phase space irregularities into a coarse grained
distribution, and neutrinos can not increase their degeneracy.

In the conventional description neutrinos are estimated insensitive to
quantum effects. \citet[p.~445]{peebles93} argues that their de
Broglie wavelength
\begin{equation}
\lambda_{\rm dB} = {h \over m_\nu v} \ ,
\end{equation}
is a small fraction of cm. The wavelength derived by Peebles,
0.001\,cm, is obtained supposing a velocity of $v = 300\, \rm km \,
s^{-1}$ and a rather large neutrino mass of $m_\nu = 123 \,\rm eV$.
Peebles' concludes: \textit{``classical physics gives an excellent
approximation to the motion of neutrinos, and Liouville's theorem
assures us the classical orbits will not violate the exclusion
principle, because the occupation number [...] is conserved''}.  The
picture is thus that neutrinos are point masses regularly distributed
in phase space, keeping approximately their relative phase space
distances sufficiently large to allow neglecting Pauli's exclusion
principle. A uniformly expanding flow of particles in space is a
uniformly shearing flow in phase space. Nearby particles in phase
space do see fluctuating neighbour distances. Therefore
if any effect depends on phase space relative distances, phase space
density perturbations can propagate.

Peebles invokes Pauli's exclusion principle and the de Broglie
wavelength as criteria for neglecting or not quantum effects.
Below we will see that relativistic and non-relativistic neutrinos are
in fact rather degenerate from the beginning, so Pauli's principle
matters.

%_______________________________________________________________________

\section{The quantum description}

\subsection{State of cosmological  neutrinos}

Let us check Peebles' assumption with recent estimates of the
neutrino properties \citep{dolgov02}, keeping in mind that the
classical world is only an approximation of the quantum world, and
that the important point is not the ratio of scales (such as the de
Broglie wavelength vs.\ the size of cosmic structures) but the ratio
of the phase space occupation density to Planck's constant $h$
cubed.

In order to not complicate the discussion, we ignore here the effect
of neutrino oscillations for which the proper collective treatment has
been only recently worked out \citep{strack05}.  It suffices here to
point out that instead of discussing the neutrino flavours (the
electron, muon and tau neutrinos which are superpositions of mass
states), we need to consider the pure mass states ($m_1$, $m_2$, and
$m_3$).  The oscillations introduce coupling terms between the
different mass states \citep{strack05}, and therefore increase the
coupling between neutrinos species.

Today's number density of electron neutrinos and anti-neutrinos
issued from the Big Bang \citep[p.~163]{dolgov02, peebles93}
of a single family is $3/11$ the number of cosmic background photons:
\begin{equation}
n_{\nu,0} = {3 \over 11} \, n_\gamma \approx 112\, {\rm cm^{-3}} \, .
\end{equation}
Therefore the mean inter-particle distance $n_{\nu,0}^{-1/3}\approx
0.2\,\rm cm$ is comparable to their de Broglie wavelength
\begin{equation}
\lambda_{\rm dB} \approx 0.4\, {\rm cm}\left(v \over 1000\,
{\rm km \, s^{-1}}\right)^{-1}\left(m_\nu \over 0.1 \,{\rm eV}\right)^{-1} \ .
\end{equation}
So with sub-eV rest mass neutrinos appear rather degenerate.  Crucial
is obviously the value of the assumed velocity. The chosen $v$ is
reasonable for neutrinos bound inside galaxy clusters, but it is not
obvious whether neutrinos could ever be trapped inside clusters.  So
let us consider the {least favourable} hypothesis that identical
neutrinos are still homogeneously distributed, with a density of 112
or 56\,cm$^{-3}$ depending on the still unknown Majorana
(anti-neutrinos are identical to neutrinos) or Dirac (anti-neutrinos
are distinct from neutrinos) nature.

%%%%%%%%%%%%%%%%%%%%%%%%%%%%%%%%%%%%%%%%%%%%%%%%%%%%%%%%%%%%%%%%%%%%
%\begin{figure}
\begin{figure*}
\sidecaption
\includegraphics[width=13.5cm]{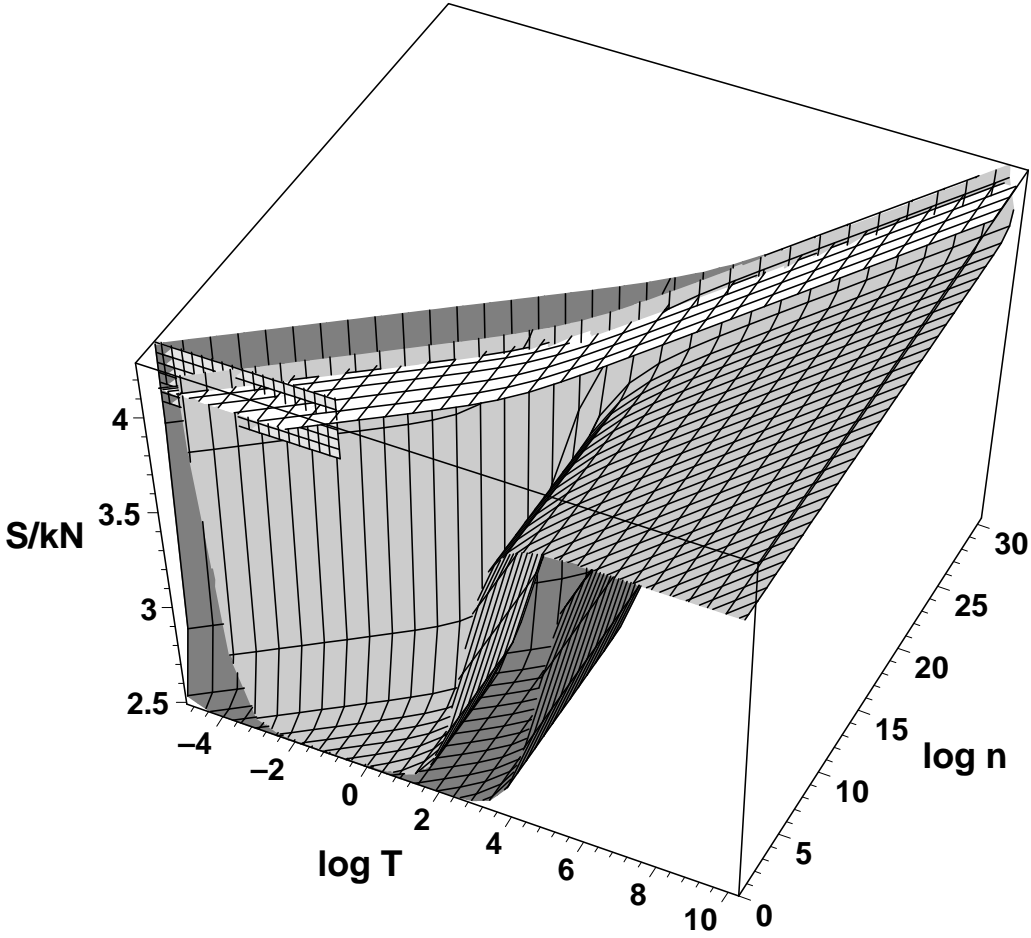}
\caption{The grey surfaces show the entropy per fermion $S/kN$ as a
  function of density and temperature ($\log n$, $\log T$) for a
  neutrino mass of $m_{\nu}=0.1\,\rm eV$ in light grey, and of
  $m_{\nu}=10\,\rm eV$ in dark grey. The horizontal white band marks
  the constant entropy of 4.2018 expected from the hot Big Bang
  fermions.  The vertical rectangle on the left shows the constant
  density of $n=56\,\rm cm^{-3}$ expected for each neutrino state at
  $z=0$.  The grid lines on the entropy surfaces run parallel to
  either constant temperature or fugacity.}
\label{fig:entropy}
%\end{figure}
\end{figure*}
%%%%%%%%%%%%%%%%%%%%%%%%%%%%%%%%%%%%%%%%%%%%%%%%%%%%%%%%%%%%%%%%%%%%%

\subsection{Entropy, fugacity, and pressure}

The accurate thermodynamical quantities for fermions at all regimes
can be calculated by evaluating numerically the relativistic
Fermi-Dirac integrals for particle density $n$, energy density $e$,
and pressure $P$ \citep[p. 216-217]{pad00}
\begin{eqnarray}
n(T,Z) &=& \frac{4\pi g_s}{h^3} \int_0^\infty\!\frac{p^2}
{Z^{-1}\exp(\epsilon/kT)+1} \, \d p
\label{equ:thermon}
\\
e(T,Z) &=& \frac{4\pi g_s}{h^3}
\int_0^\infty\!\frac{p^2\epsilon} {Z^{-1}\exp(\epsilon/kT)+1} \, \d p
\label{equ:thermoe}
\\
P(T,Z) &=& \frac{4\pi g_s}{h^3} \int_0^\infty\!\frac{p^2}
{Z^{-1}\exp(\epsilon/kT)+1}\,\frac{1}{3}\,\frac{c^2
p^2}{\epsilon+mc^2} \, \d p
\label{equ:thermop}
\end{eqnarray}
where $g_s$ is the number of distinct particle states, $\epsilon=
\sqrt{p^2c^2+m^2c^4} -m c^2$ is the particle energy, and $Z$ is the
fugacity (the degeneracy parameter), capitalised here to distinguish
from redshift $z$.  The calculations in this paper have been performed
with the software package Maple, version 10, in which the number of
significant digits can be chosen.  Here 14 digit precision has been used.

The adiabatic expansion of neutrinos since redshift $z=10^{10}$ means
that the entropy per particle $S/kN$, a dimensionless number, has been
constant since then. If the initial chemical potential is negligible,
$S/kN$ amounts to 4.2018 for fermions, and 3.6016 for bosons
\citep[p. 277]{peacock99}. The entropy per particle can be calculated from
Eqs.~(\ref{equ:thermon}-\ref{equ:thermop}) with the relationship $S/kN
= (e + P)/nkT$, a function of $T$ and $Z$, which no longer depends on
$g_s$ and $h$.
%
%\begin{eqnarray}
%{S \over kN}(T,Z) &=& \int_0^\infty\!...
%\end{eqnarray}
%
Calculating the implicit function $S/kN(T,Z)$ for a sufficiently wide
range of $T$ and $Z$, we show in Fig.~\ref{fig:entropy}  its dependence on
$n$ and $T$ over relevant cosmological values from high to low
redshifts, for two representative neutrino masses, 0.1 and 10\,eV.
The grey parts, almost vertical in the back, correspond to degenerate
states, with $n \propto T^{3/2}$ on the right in the relativistic regime,
and $n \propto T^{3}$ on the left in the non-relativistic regime.
The almost horizontal parts in the front of the $S/kN$ maps correspond
to non-degenerate states not reached during the Universe expansion. The step
passing from $S/kN=2.5$ to 4 around $\log T = 2-4$ corresponds to the
transition from the non-degenerate non-relativistic regime to the
relativistic regimes. The position of the step does depend on the particle
mass. Clearly, today's cosmological neutrinos with sub-eV masses are
much colder than the frequently quoted 1.95\,K temperature valid for
relativistic neutrinos.

%%%%%%%%%%%%%%%%%%%%%%%%%%%%%%%%%%%%%%%%%%%%%%%%%%%%%%%%%%%%%%%%%%%%%%%%
\begin{figure}
\vspace{4mm}
\resizebox{\hsize}{!}{\includegraphics{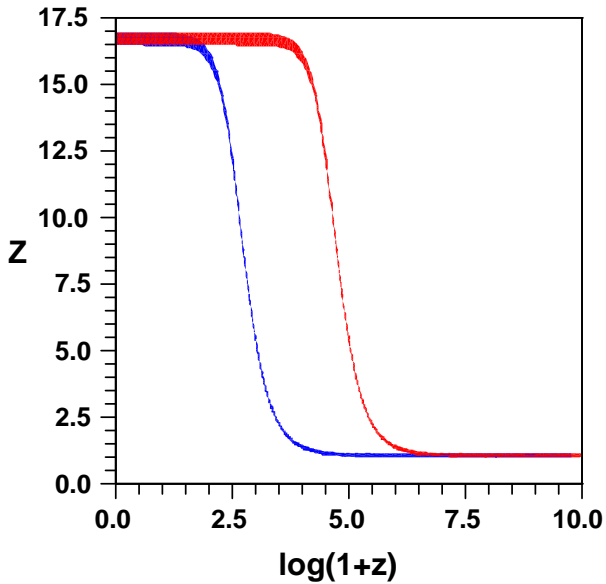}}
\vspace{-7mm}
\caption{Fugacity $Z$ as a function of redshift $z$ for constant
comoving fermion density $n=56\,(1+z)^3 \,\rm cm^{-3}$ and constant
entropy per fermion $S/kN$ in the range $4.2018$ (lower limit of the
bands) and $4.24$ (upper limit of the bands) for $m_{\nu}=0.1\,\rm eV$
(lower band), and $m_{\nu}=10\,\rm eV$ (upper band).}
\label{fig:fugacity}
\end{figure}
%%%%%%%%%%%%%%%%%%%%%%%%%%%%%%%%%%%%%%%%%%%%%%%%%%%%%%%%%%%%%%%%%%%%%%%%

Observing carefully the constant fugacity curves in
Fig.~\ref{fig:entropy}, one notices that the adiabatic expansion from
the relativistic to the non-relativistic regime \textit{increases} the
neutrino degeneracy, since the constant fugacity curves drop their
entropy level.  To make this point quantitative, we solve numerically
for $T$ and $Z$ the pair of equations
\begin{equation}
\frac{S}{kN}(T,Z)=4.2018 \ , \qquad
n(T,Z)=56 \, \rm cm^{-3}\ ,
\end{equation}
with a bisection search technique because the non-linear equation 
solver of Maple 
for floating-point number solutions \texttt{fsolve} based on 
a Newton-Raphson method fails in this case. 
The value of $n$ is a low value for today's neutrino density
(choosing $n=112\, \rm cm^{-3}$ would reinforce our conclusions). We
find that the temperature scales as the inverse of the mass, while $Z$
at $z=0$ is practically independent of the neutrino mass.  We obtain
for the present neutrino temperature,
\begin{equation}
T_{\nu,0} =  1.614\cdot 10^{-4}\, \left( \frac{m_\nu}{1\rm
eV}\right)^{-1} {\rm K}, \qquad {\rm and}, \qquad Z = 16.50.
\label{equ:temp}
\end{equation}
So between the creation of neutrinos at $z \approx 10^{10}$ and now,
somehow fugacity has increased from 1 to 16.5.

To show how fugacity behaves as function of redshift at constant
entropy per particle, we ask Maple to solve numerically by 
interpolation the above
quantities in Eqs.~(\ref{equ:thermon}-\ref{equ:thermop}).
The result is shown in Fig.~\ref{fig:fugacity}.  Fugacity increases
from the early Universe to the present by a factor 16.5: massive
neutrinos are substantially more degenerate now than at their
decoupling epoch.  This is independent of their mass if they are
non-relativistic.  When calculating the $Z(z)$ relationship for a
range of entropies, we see that if for any reason entropy would
increase instead of being strictly constant, fugacity would
\textit{increase} even more.

%%%%%%%%%%%%%%%%%%%%%%%%%%%%%%%%%%%%%%%%%%%%%%%%%%%%%%%%%%%%%%%%%%%%%%%%
\begin{figure}
\resizebox{\hsize}{!}{\includegraphics{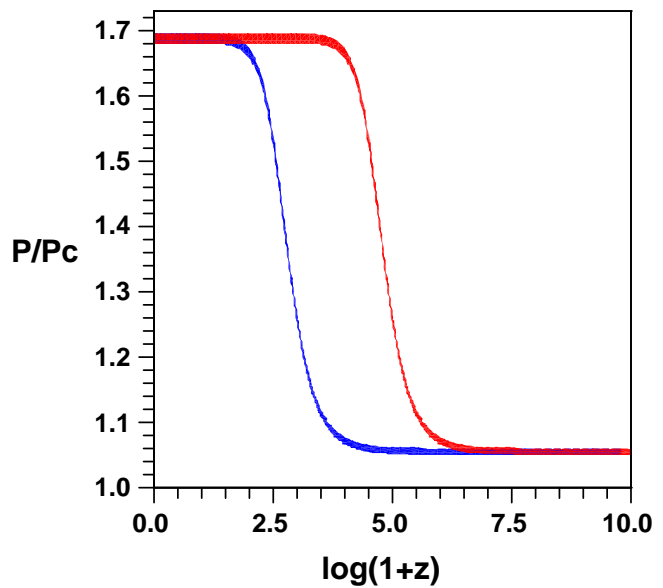}}
\caption[]{Ratio of neutrino pressure $P$ to a perfect gas pressure
  $P_c= nkT$, as a function of redshift $z$ for constant comoving
  fermion density $n=56\,(1+z)^3 \,\rm cm^{-3}$ and constant entropy
  per fermion $S/kN$ in the range $4.2018$ (lower limit of the bands)
  and $4.24$ (upper limit of the bands) for $m_{\nu}=0.1\,\rm eV$
  (lower band), and $m_{\nu}=10\,\rm eV$ (upper band).}
\label{fig:pressure}
\end{figure}
%%%%%%%%%%%%%%%%%%%%%%%%%%%%%%%%%%%%%%%%%%%%%%%%%%%%%%%%%%%%%%%%%%%%%%%%

Associated to fermion degeneracy is the Fermi pressure contribution.
To quantify this important factor for dynamics, again we ask Maple to
evaluate numerically the integrals in
Eqs.~(\ref{equ:thermon}-\ref{equ:thermop}), and to find by
interpolation the pressure at constant specific entropy curve.  In
Fig.~\ref{fig:pressure} the ratio of the full pressure to the perfect
gas pressure  $P_c= nkT$ is displayed.  The additional quantum pressure with
respect to a collisionless gas amounts to 5\% at high redshifts,
independent of the neutrino mass.  At the transition from the
relativistic to the non-relativistic regime, the redshift of which
does depend on the mass, the quantum pressure contribution increases
to 68\%, and becomes subsequently independent on the mass in the
non-relativistic regime.  Again any increase in specific entropy would
increase pressure.

Obviously this additional pressure is a substantial effect at low
redshifts determining the clustering of neutrinos during the structure
formation phase.  The immediate prediction is that neutrinos should
tend to accumulate less in bound structures than truly collisionless
particles.  However, dissipative effects in non-linearities such as
shocks may eventually falsify this prediction.

\subsection{Asymptotic estimates}

These previous results are exact down to the used 14-digit numerical
precision. Let us now also derive intermediate states with asymptotic
approximations, in part to estimate the errors made with this
more traditional approach.

During the relativistic phase of expansion neutrinos cool
proportionally to the photon temperature \citep{peebles93}:
\begin{equation}
T_{\nu,r} = \left(4 \over 11\right)^{1/3} T_\gamma = \left(4 \over
11\right)^{1/3} T_0\, (1+z) \, ,
\end{equation}
where $T_0$ is today's CMB temperature of 2.73\,K.  When the neutrino
thermal energy equals its rest mass energy, neutrinos become
non-relativistic.  Then
\begin{equation}
m_{\nu,r} c^2 = 3 k T_\nu \approx \left(4 \over 11\right)^{1/3} 3 k
T_0 \,(1+z) \, .
\end{equation}
Then the critical temperature is
\begin{equation}
  T_{\nu,c} = { m_\nu c^2 \over 3k } \approx 3970 \left( m_\nu \over
  {\rm 1\, eV} \right)\, {\rm K}\ ,
\end{equation}
and the critical redshift
\begin{equation}
  1+z_c = 1986 \left( m_\nu \over {\rm 1\, eV} \right) \, .
\end{equation}
Below redshift $z_c$ temperature drops for adiabatic non-relativistic
matter approximately as
\begin{equation}
   T_{\nu,n} = T_{\nu,c} \left( 1+z \over 1+z_c \right)^2 \approx 9.8
             \cdot 10^{-4}\, (1+z)^2 \left( m_\nu \over {\rm 1 \,eV}
             \right)^{-1} \, {\rm K} \, .
\end{equation}
Thus the thermal speed of cosmological neutrinos amounts to
\begin{equation}
   v_\nu = \left(4\over 11\right)^{1/3} {3 kT_0 \over m_\nu c} \, (1+z)
    \approx 151 \, (1+z) \, \left( m_\nu \over {\rm 1\, eV}
    \right)^{-1} \, {\rm km\, s^{-1}} \, .
\end{equation}
\citep[see][for a similar derivation]{ringwald04}.  With respect to
the exact result Eq.(\ref{equ:temp}), $T_{\nu,n}$  is
overestimated by a factor 6.07, and $v_\nu $ by a factor 2.46.

Owing to number conservation, the neutrino density $n_\nu$ varies 
in proportion of the volume
\begin{equation}
   n_\nu = n_{\nu,0} \, (1+z)^3 \, .
\end{equation}
The average phase space volume per neutrino is estimated as
\begin{equation}
   V_\nu \sim {1 \over n_\nu} {4\pi v_\nu^3 \over 3}  \approx {144 \pi
   \over 11 n_0}
   \left( k T_0 \over c\right)^3 \, ,
\end{equation}
which is independent of the neutrino mass and redshift, as long
as neutrinos are non-relativistic.  Compared with the elementary
quantum volume $h^3$, we have,
\begin{equation}
   {V_\nu \over h^3} \approx {284 \over n_{\nu,0}} \, \approx 5.1
\left( n_{\nu,0} \over 56 \right)^{-1}\, .
\end{equation}
Neutrinos appear therefore rather close to be strongly degenerate,
consistently with the previous accurate calculations.
Therefore quantum effects must be considered with more attention.

%_______________________________________________________________________

\section{Quantum Effects}

\subsection{The paradox of collisionless identical
particles in the quantum regime}

In view of the previous results, we are led to a paradox. On one side
from cosmological conditions we find that neutrinos are fairly densely
packed in phase space, so Pauli's exclusion principle should be
applicable, while on the other hand in the usual view among
cosmologists neutrinos should ignore each other since their
particle-particle electro-weak coupling is very low. How can
particles without interaction term interact?

Often in quantum mechanics one must use the idealisations of isolated,
independent systems, although wave-functions formally extend over the
whole space.  This is justified 
\citep[e.g.,][chap.~XIV, D, p.~1384]{cohen00} 
when the system wave-function is itself well separated from
the outer world wave-functions. This is possible because particles
are emitted not as plane waves extending uniformly over the whole
space, but as wave-packets since any emission process takes a positive
time-interval. Therefore the individual wave-functions consist of
well defined square integrable functions with compact support, as
required by quantum theory.

So if we picture individual neutrinos as localised moving
wave-packets, they follow a classical free particle trajectory as
long as the wave-packets are well separated. But when the wave-packet
cores of identical particles overlap, then quantum interference must
occur, blurring the possibility to distinguish the particles, because
they must simultaneously follow Pauli's rule of anti-symmetrisation for
fermions. Pauli's symmetrisation rules for fermions and bosons are
independent of the particular particle kind and strength of the
interaction potential. 

All the interactions between elementary particle, nuclei, atoms,
molecules, and even larger pieces of matter like superfluids, follow
the Pauli symmetrisation rules: whenever particle wave-functions
overlap, quantum mechanics is consistent with experiments at the
condition of applying the wave-function symmetry postulate.  Somehow
the individual packets of identical particles which could be
considered as independent when far apart, must form a symmetric or
antisymmetric combined wave-function when they overlap.  The speed of
this (anti-)symmetrisation process has nothing to do with the
electro-weak interaction, but just the quantum mechanical nature of
identical particles.

Before discussing this paradox further, let us illustrate graphically
with exact solutions of the \Sch equation how the wave-packets of two
identical free fermions behave.

%%%%%%%%%%%%%%%%%%%%%%%%%%%%%%%%%%%%%%%%%%%%%%%%%%%%%%%%%%%%%%%%%%%%%%%%
\begin{figure*}
\sidecaption
{
  \includegraphics[width=6cm]{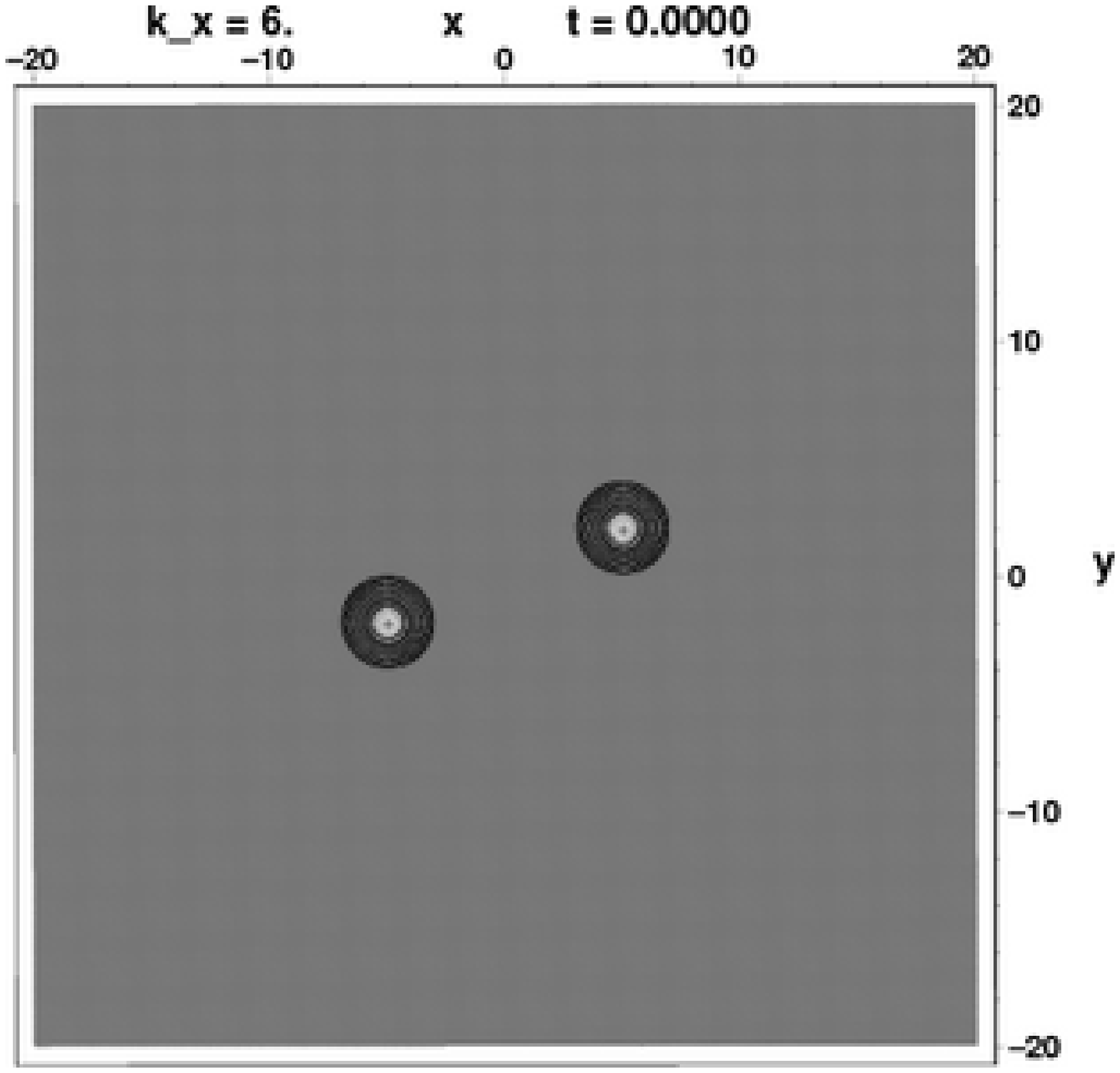}
  \includegraphics[width=6cm]{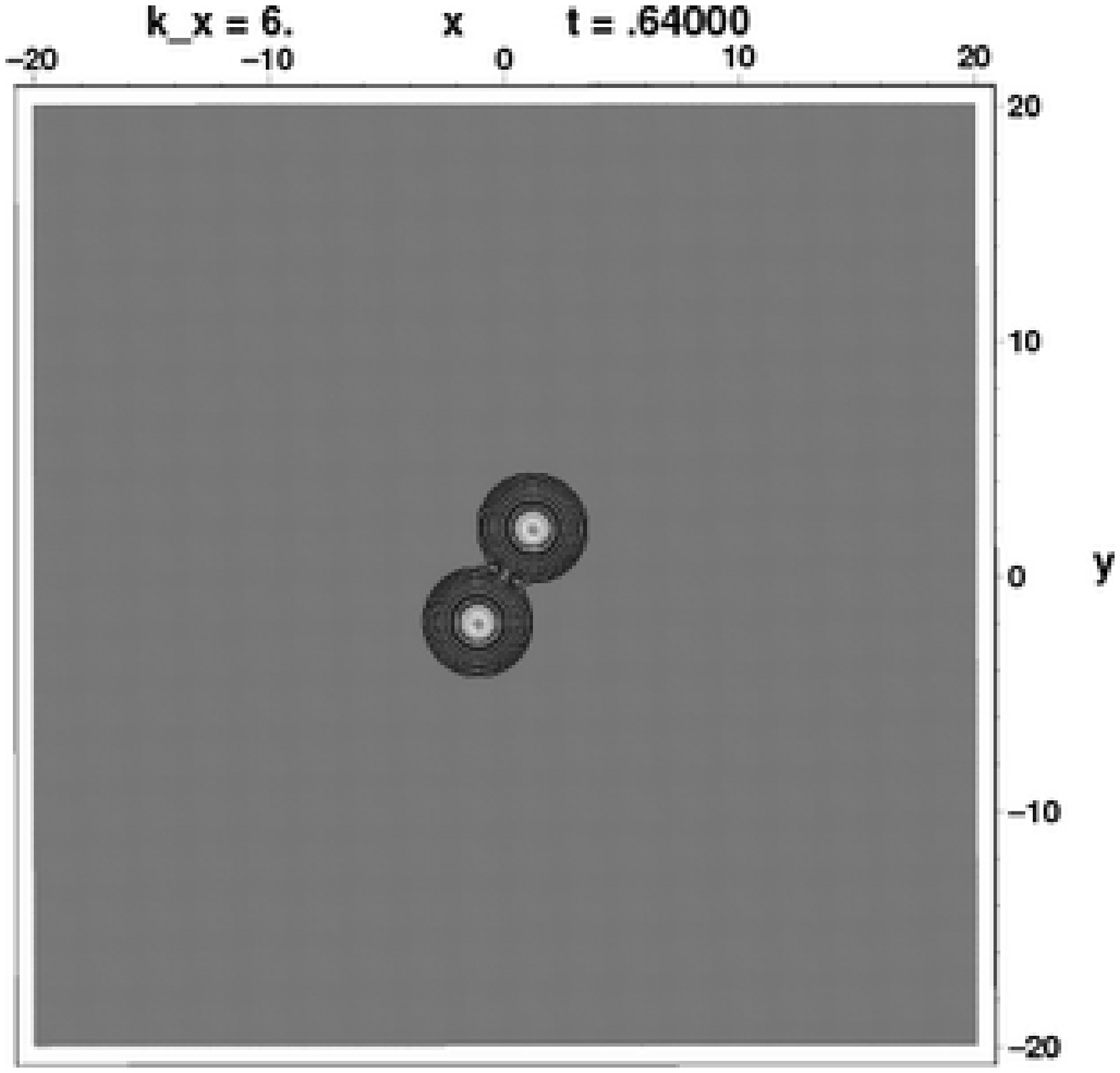}
\bigskip
  \includegraphics[width=6cm]{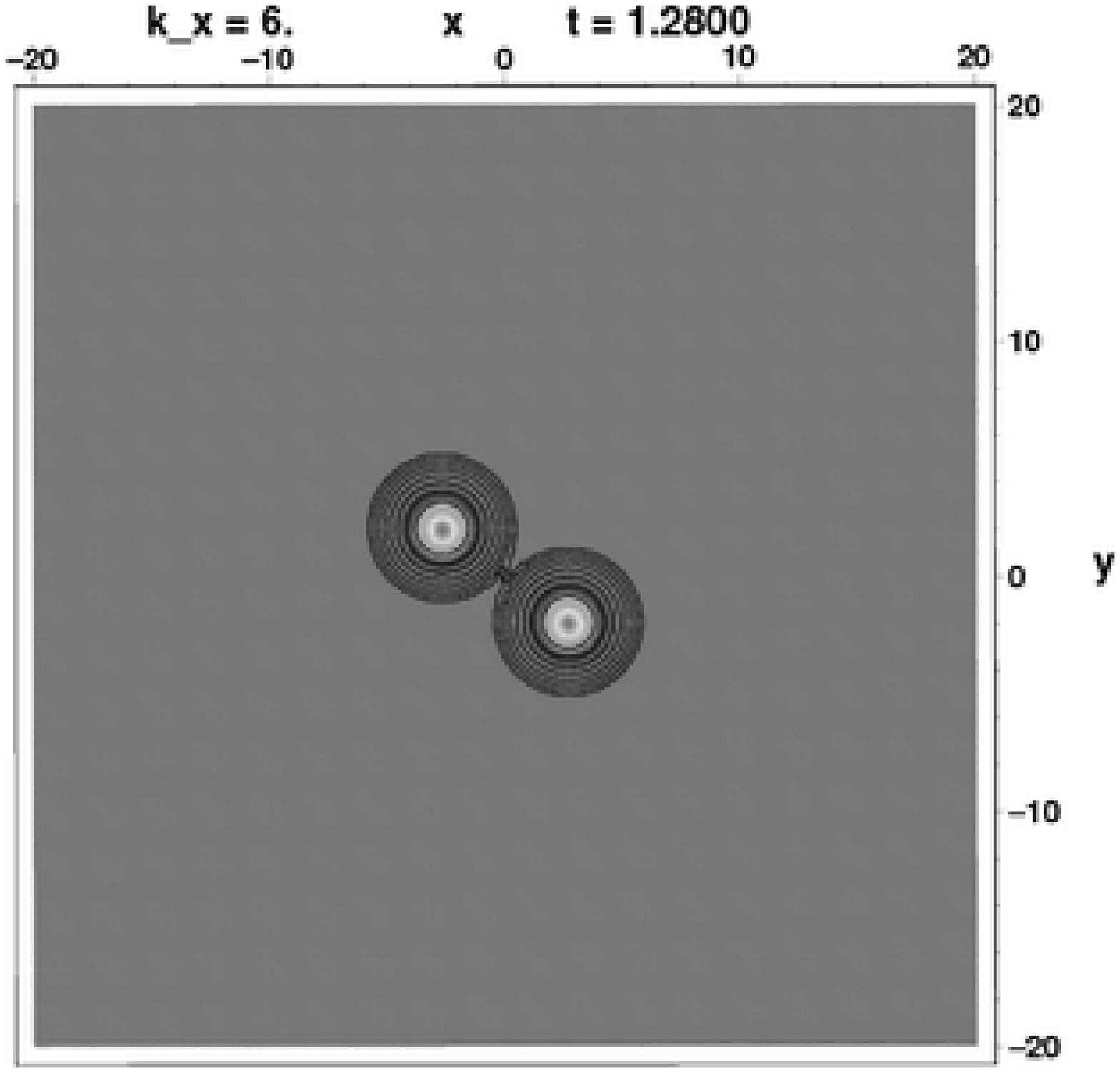}
  \includegraphics[width=6cm]{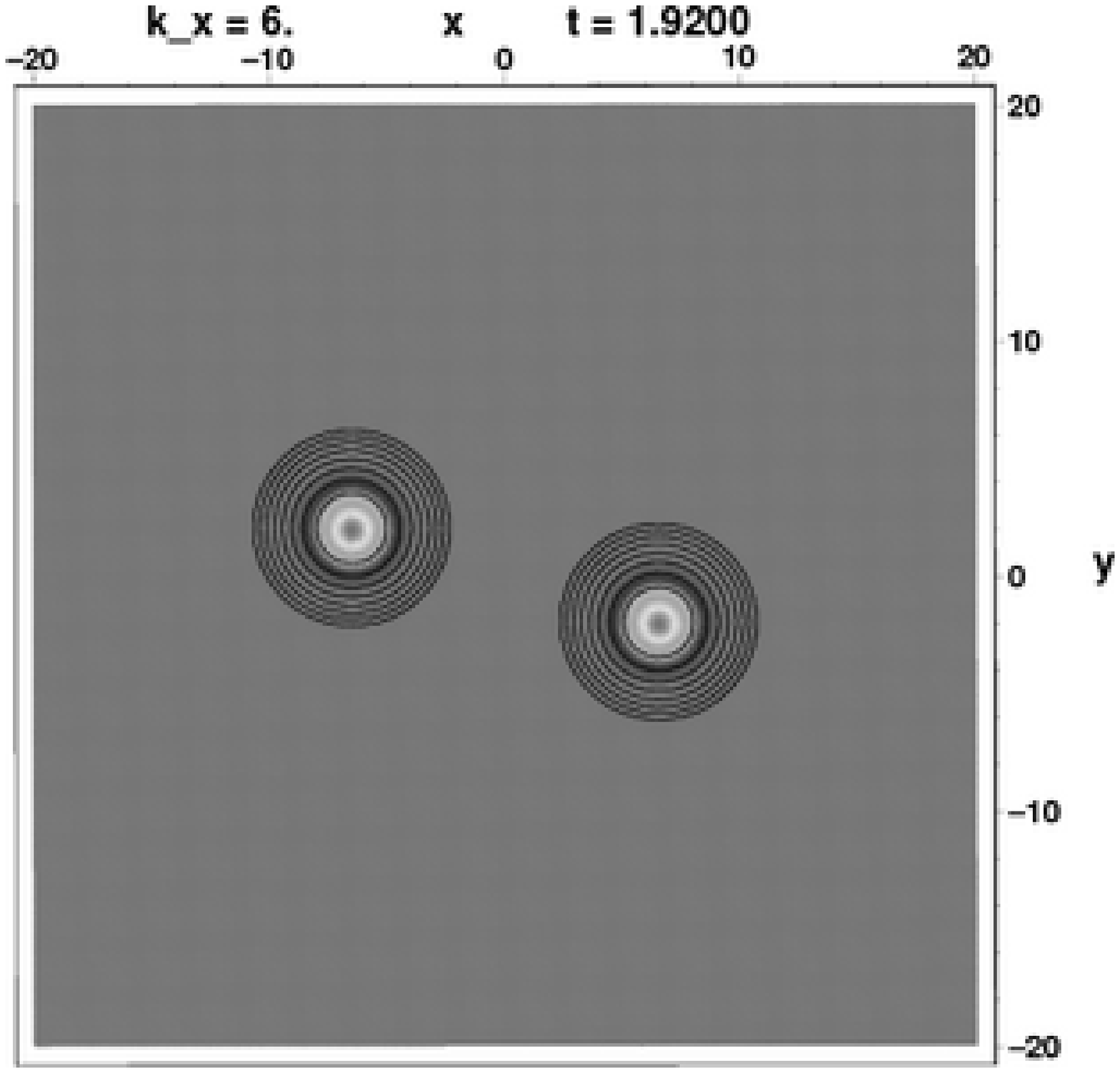}
  \includegraphics[width=6cm]{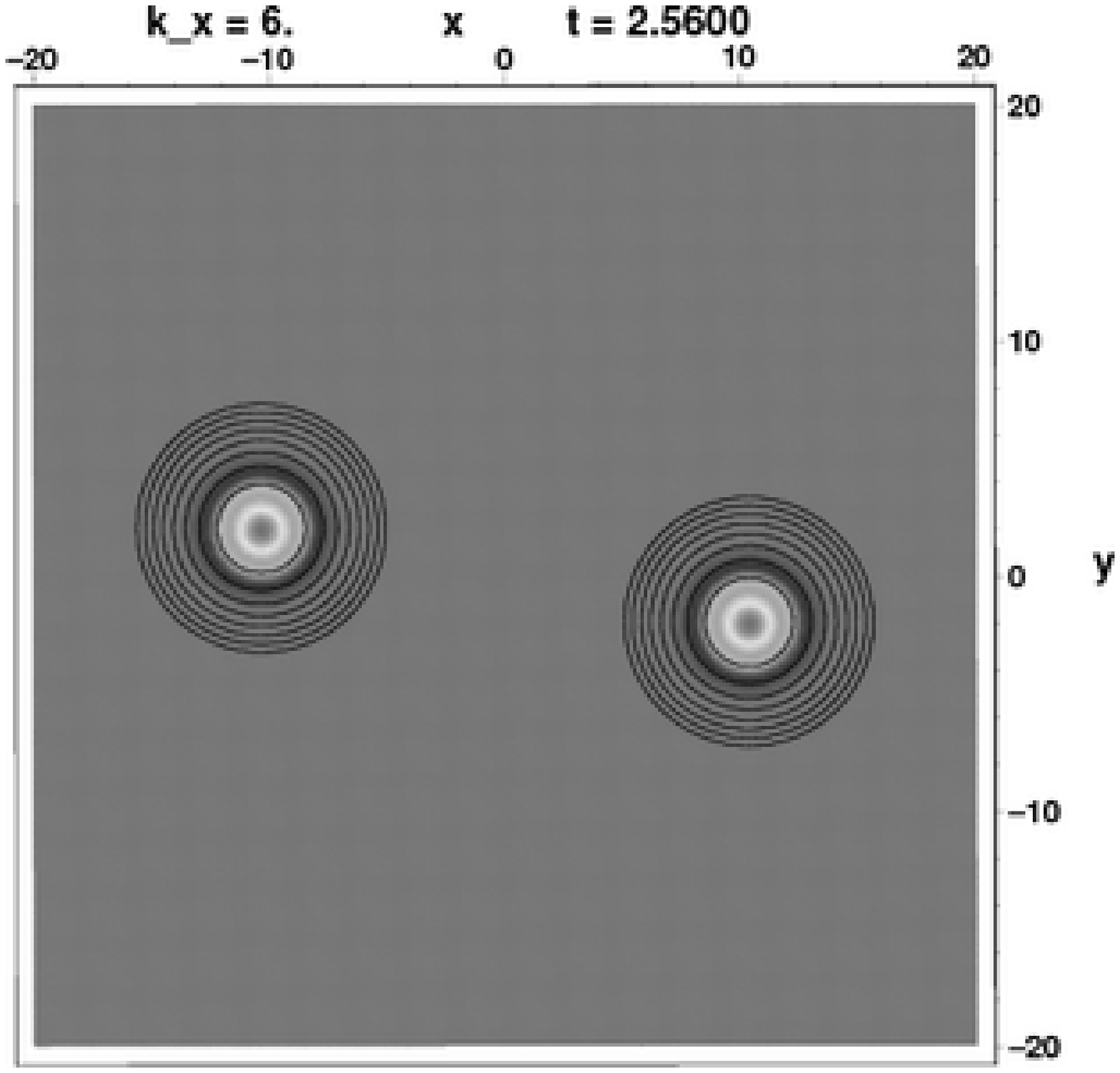}
  \includegraphics[width=6cm]{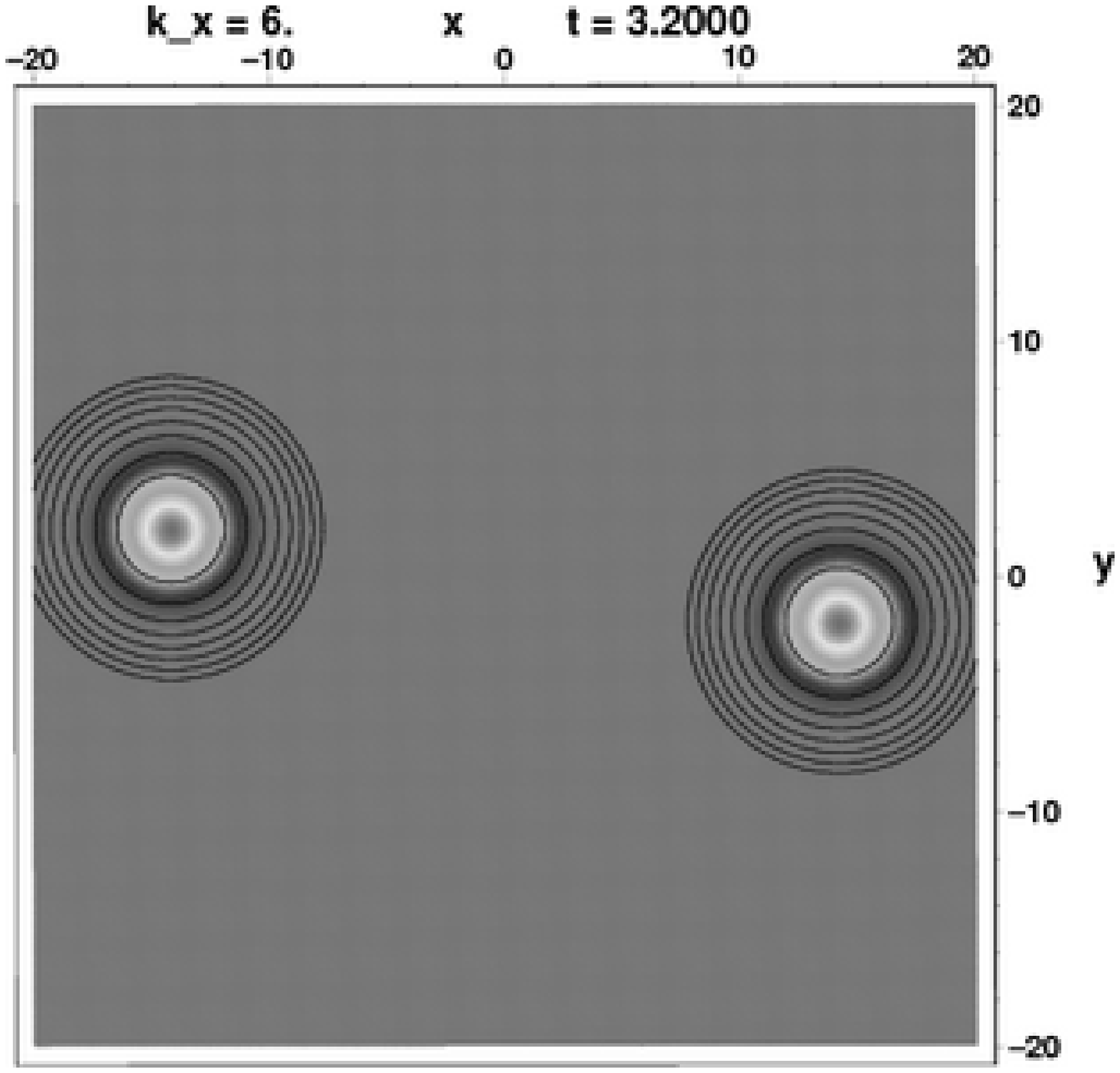}
}
  \caption{Interaction of two free fermions in the classical regime
    starting initially in opposite directions and horizontally
    ($k_x=6$, $k_y=0$, $k_z=0$).  Shown is the normalised squared norm
    $|\Psi_-(\vec x,-\vec x,t)|^2$ in the plane $(x,y,0)$. }
  \label{fig:classical}
\end{figure*}

%%%%%%%%%%%%%%%%%%%%%%%%%%%%%%%%%%%%%%%%%%%%%%%%%%%%%%%%%%%%%%%%%%%%%%%%

\begin{figure*}
\sidecaption {
\includegraphics[width=6cm]{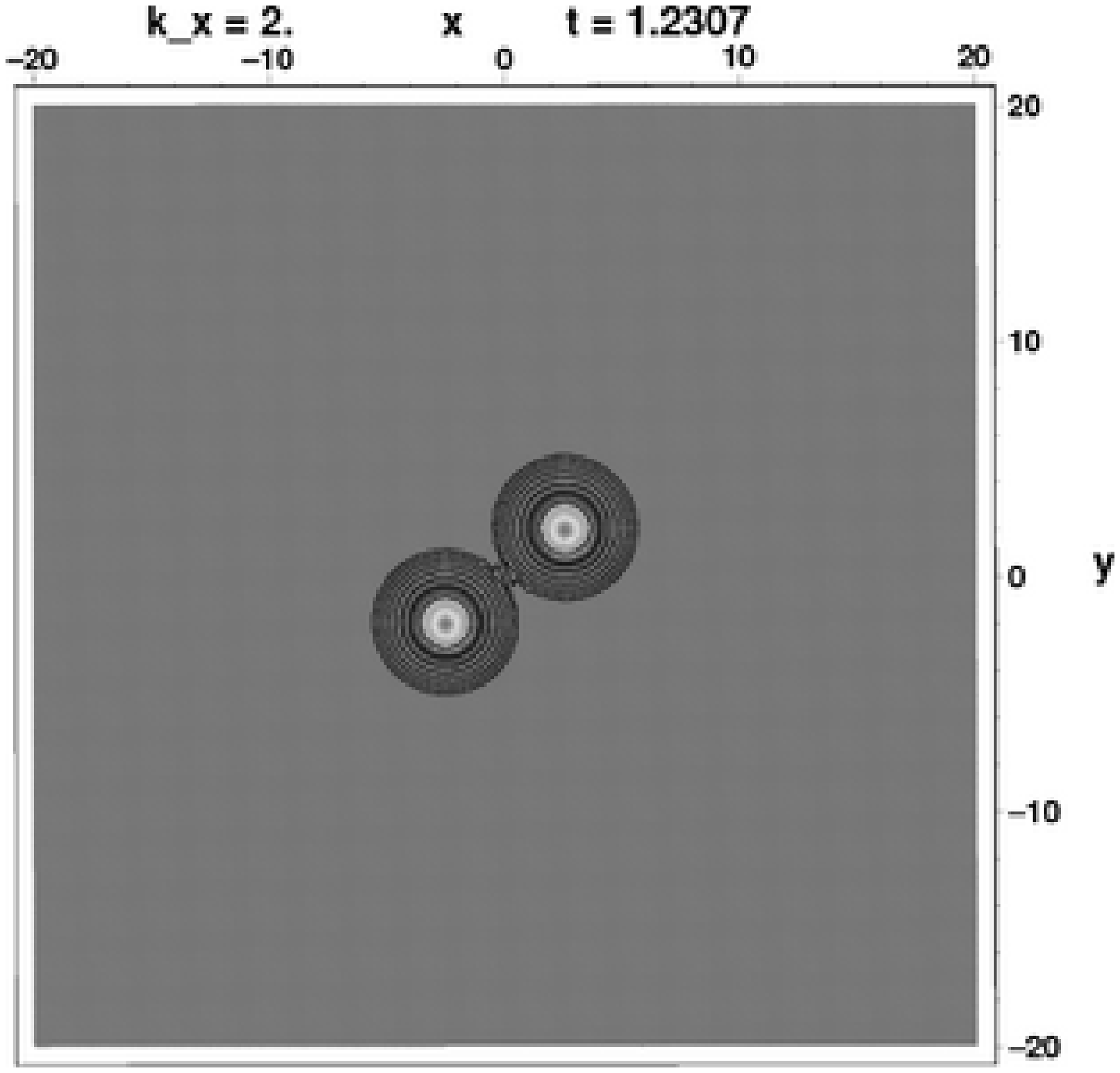}
\includegraphics[width=6cm]{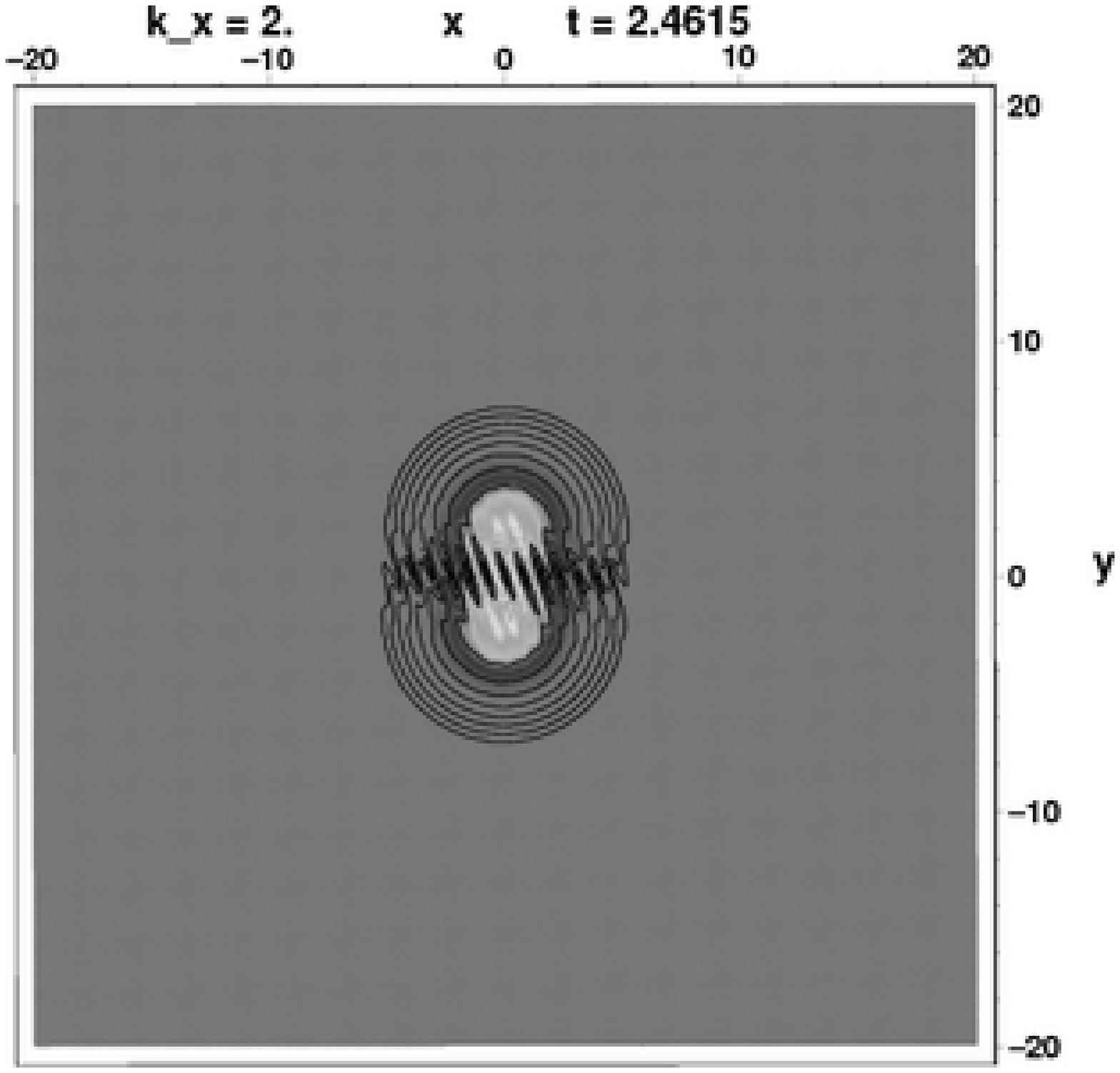}
\bigskip
\includegraphics[width=6cm]{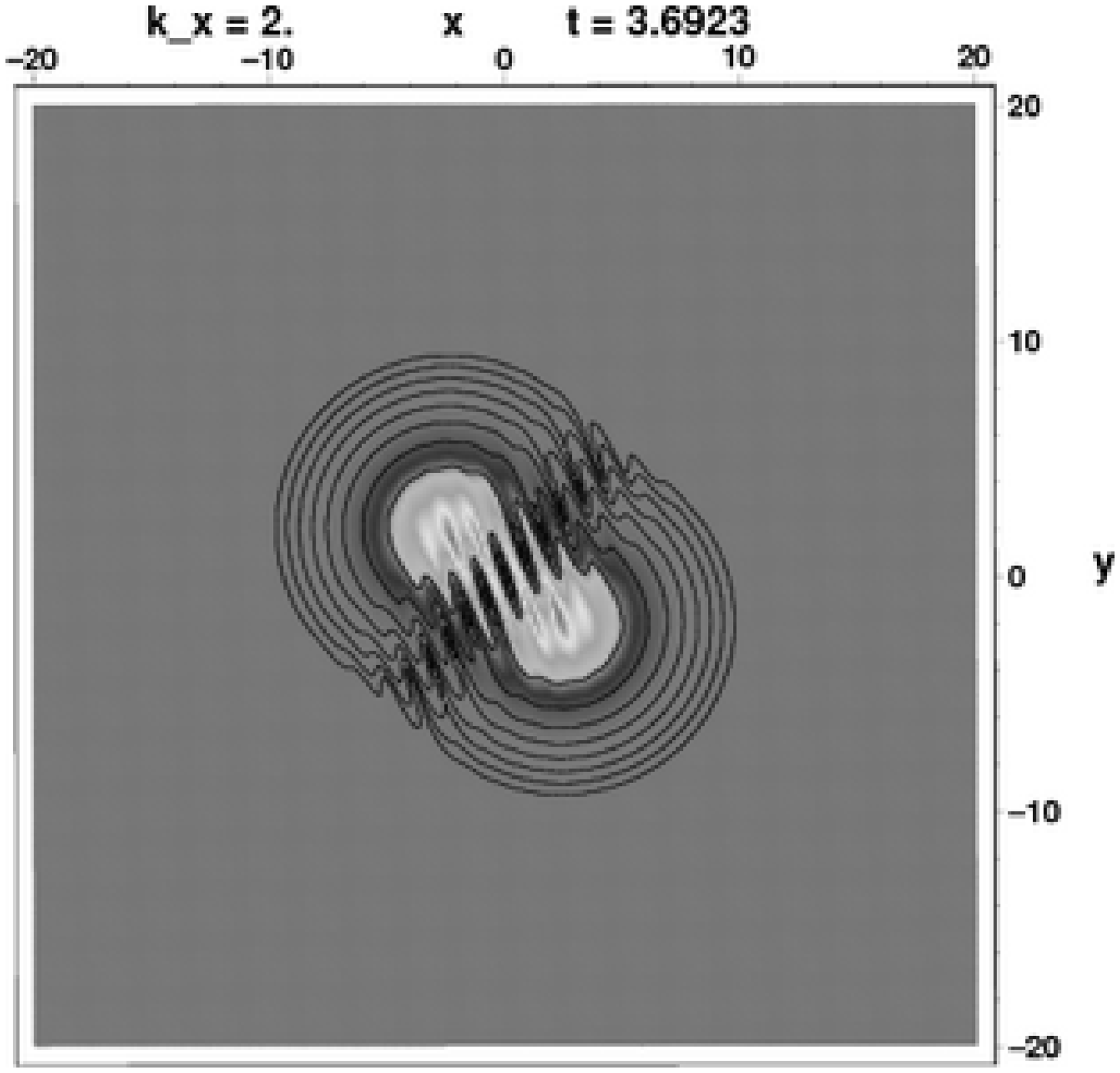}
\includegraphics[width=6cm]{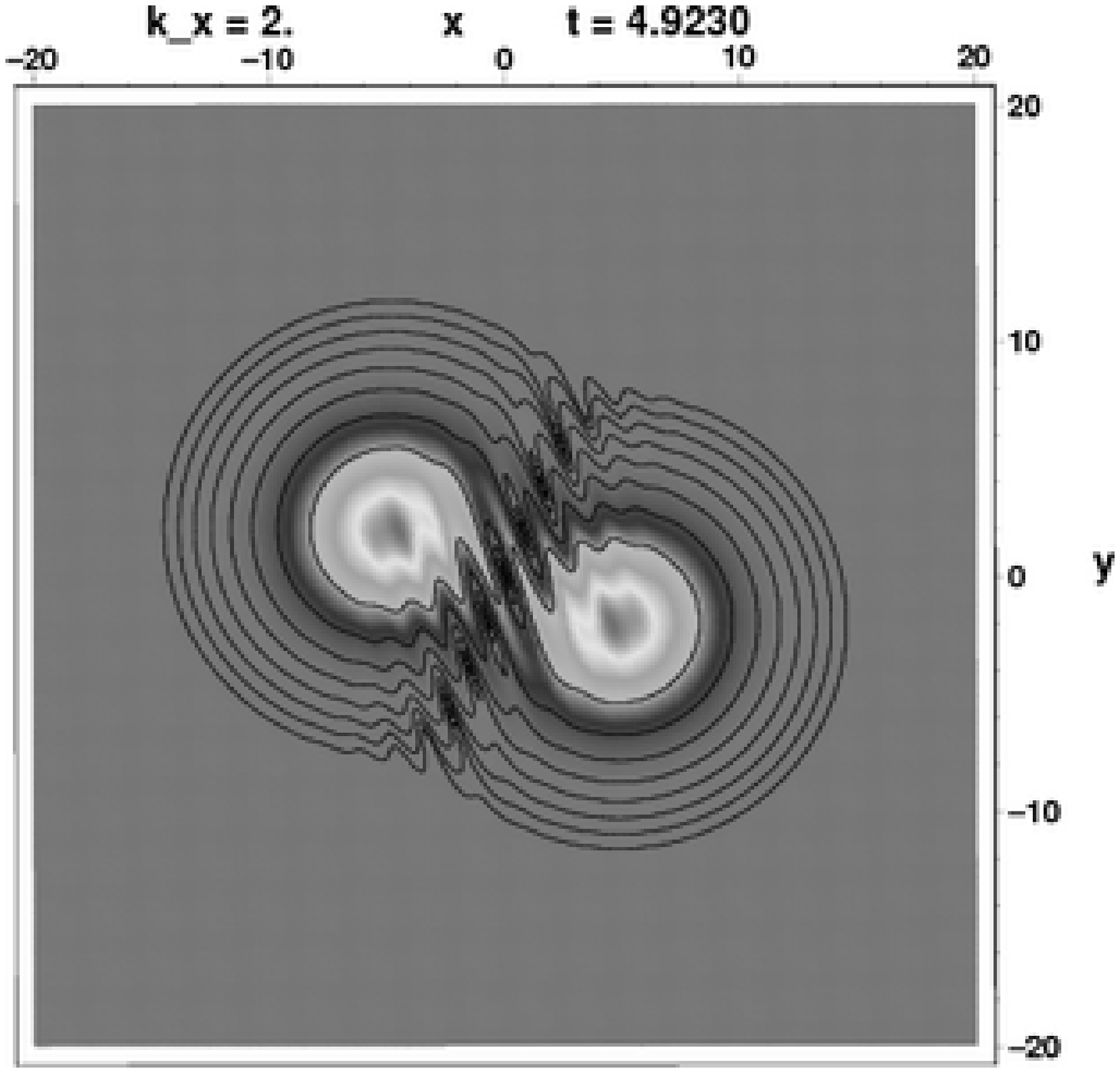}
\includegraphics[width=6cm]{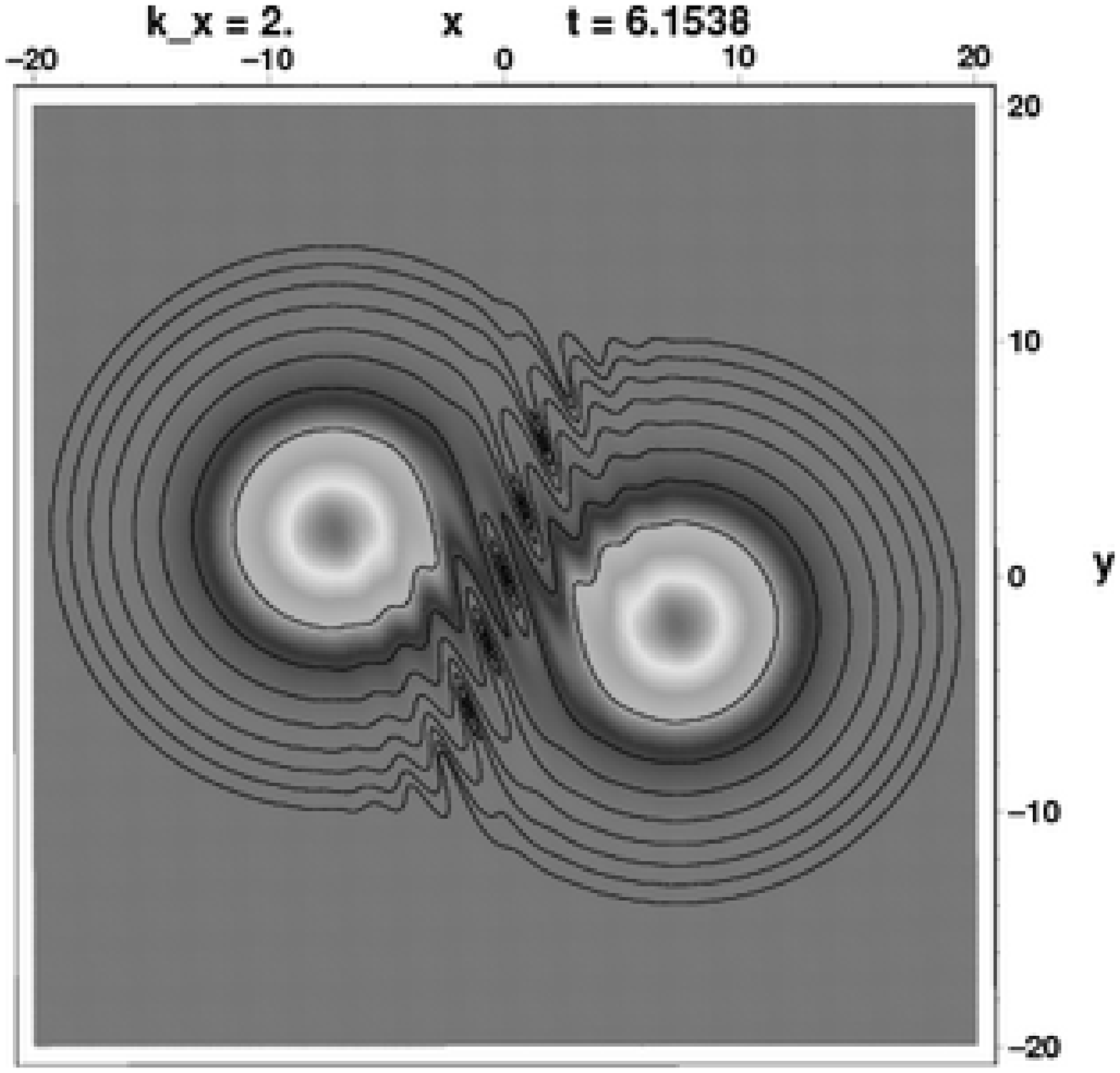}
\includegraphics[width=6cm]{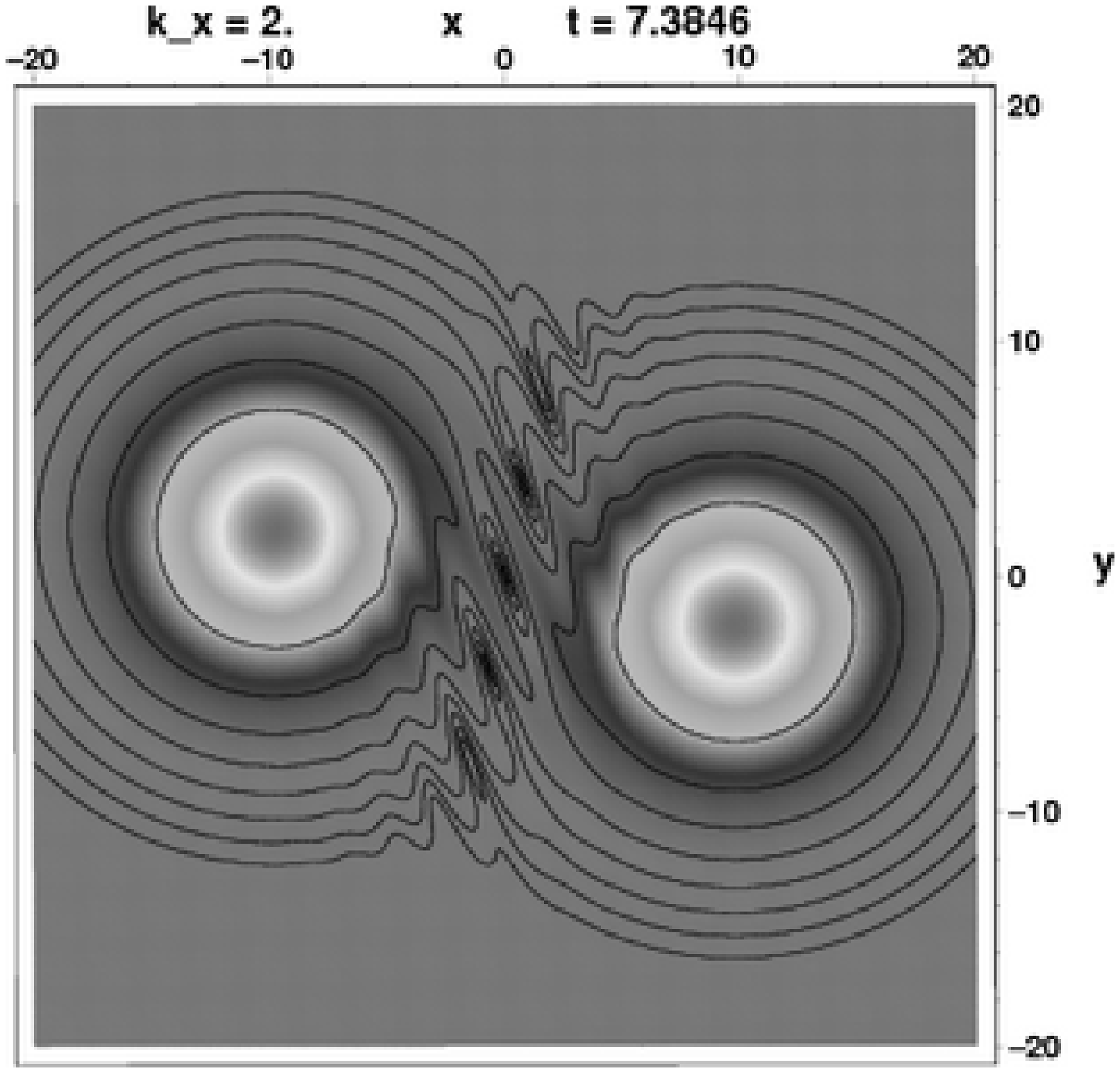}
}
   \caption{As in Fig.~(\ref{fig:classical}), but in the
   semi-classical regime ($k_x=2$)}
   \label{fig:semi-classical}
\end{figure*}

%%%%%%%%%%%%%%%%%%%%%%%%%%%%%%%%%%%%%%%%%%%%%%%%%%%%%%%%%%%%%%%%%%%%%%%%

\begin{figure*}
\sidecaption
{
\includegraphics[width=6cm]{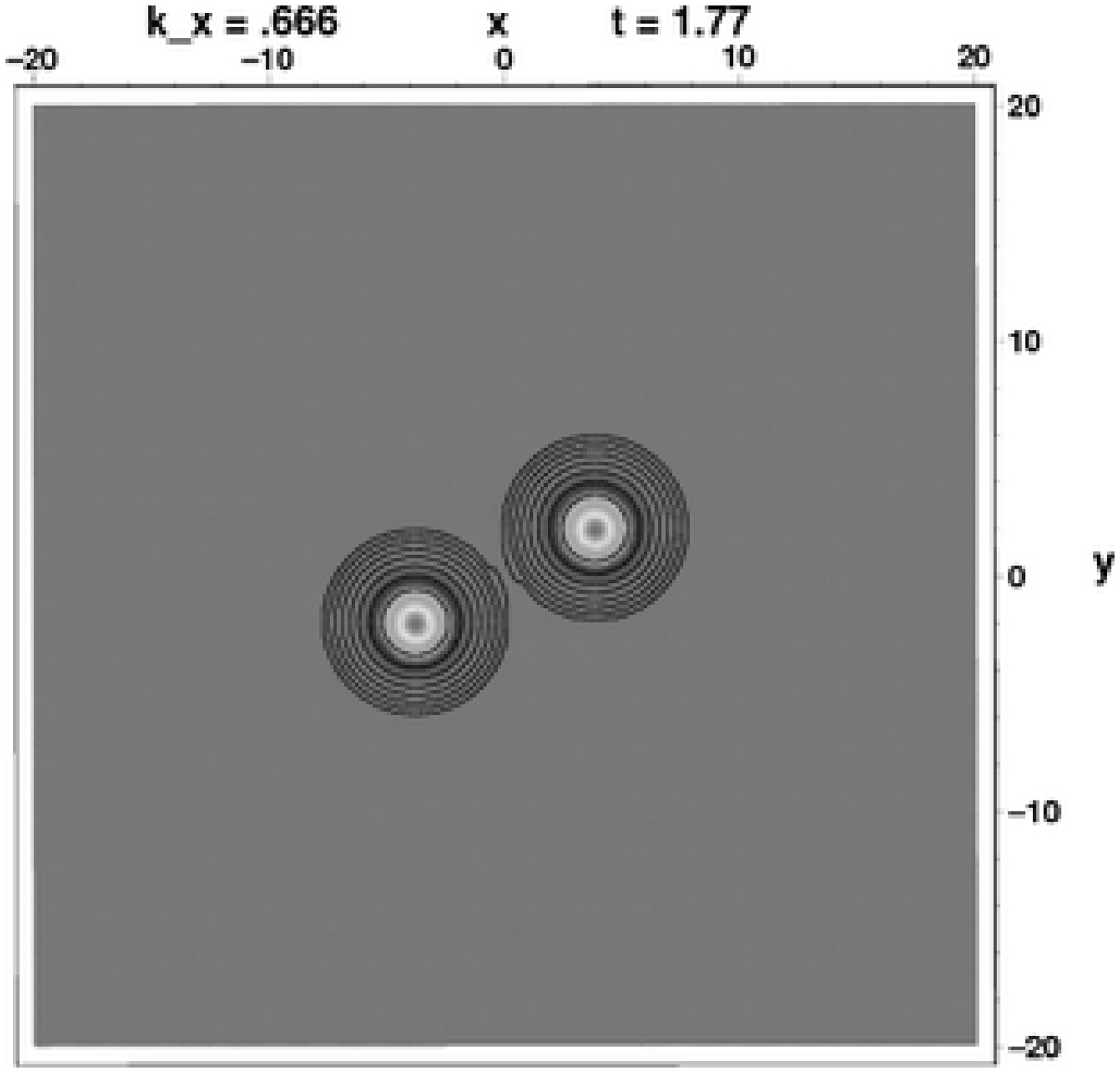}
\includegraphics[width=6cm]{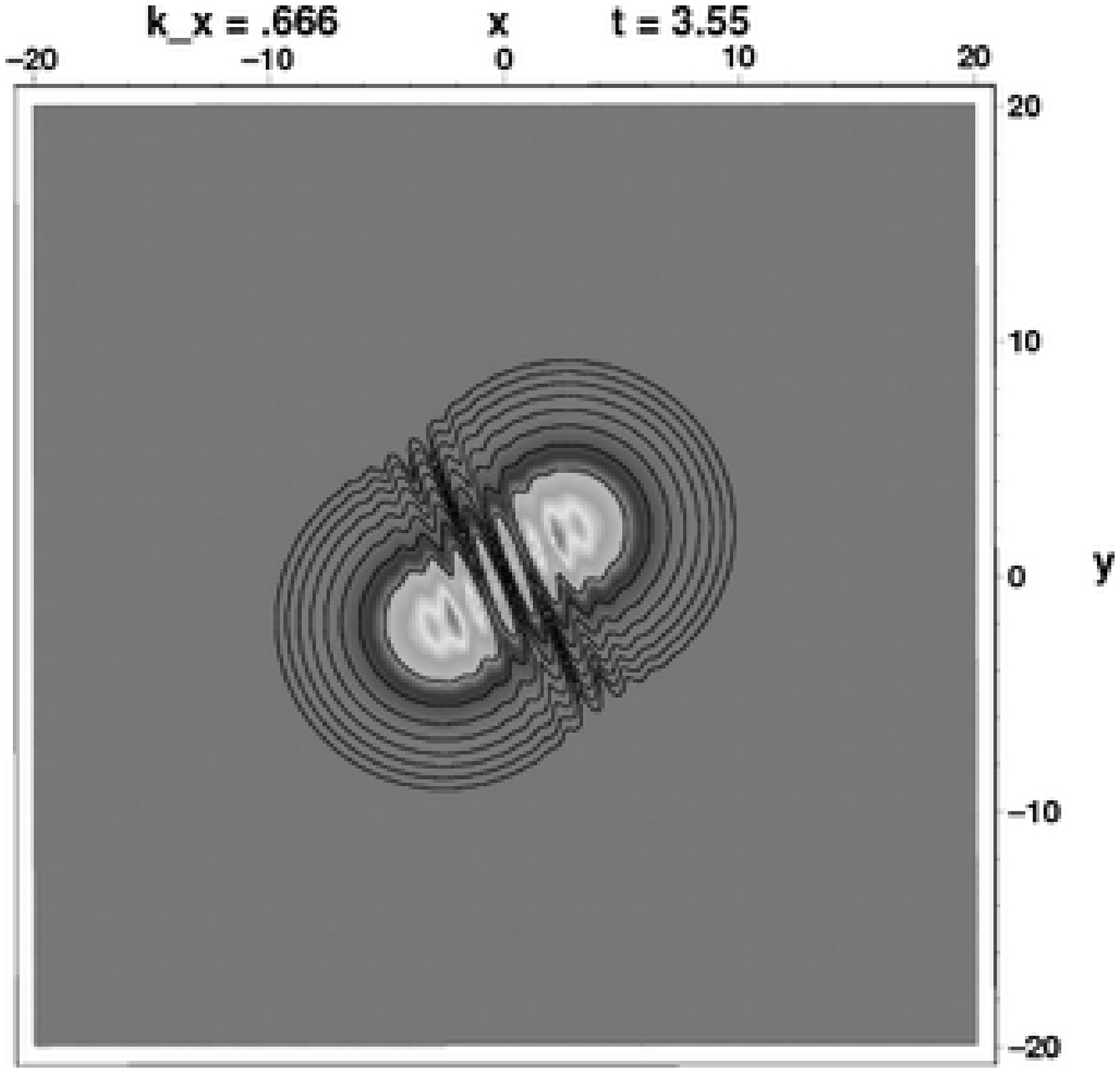}
\bigskip
\includegraphics[width=6cm]{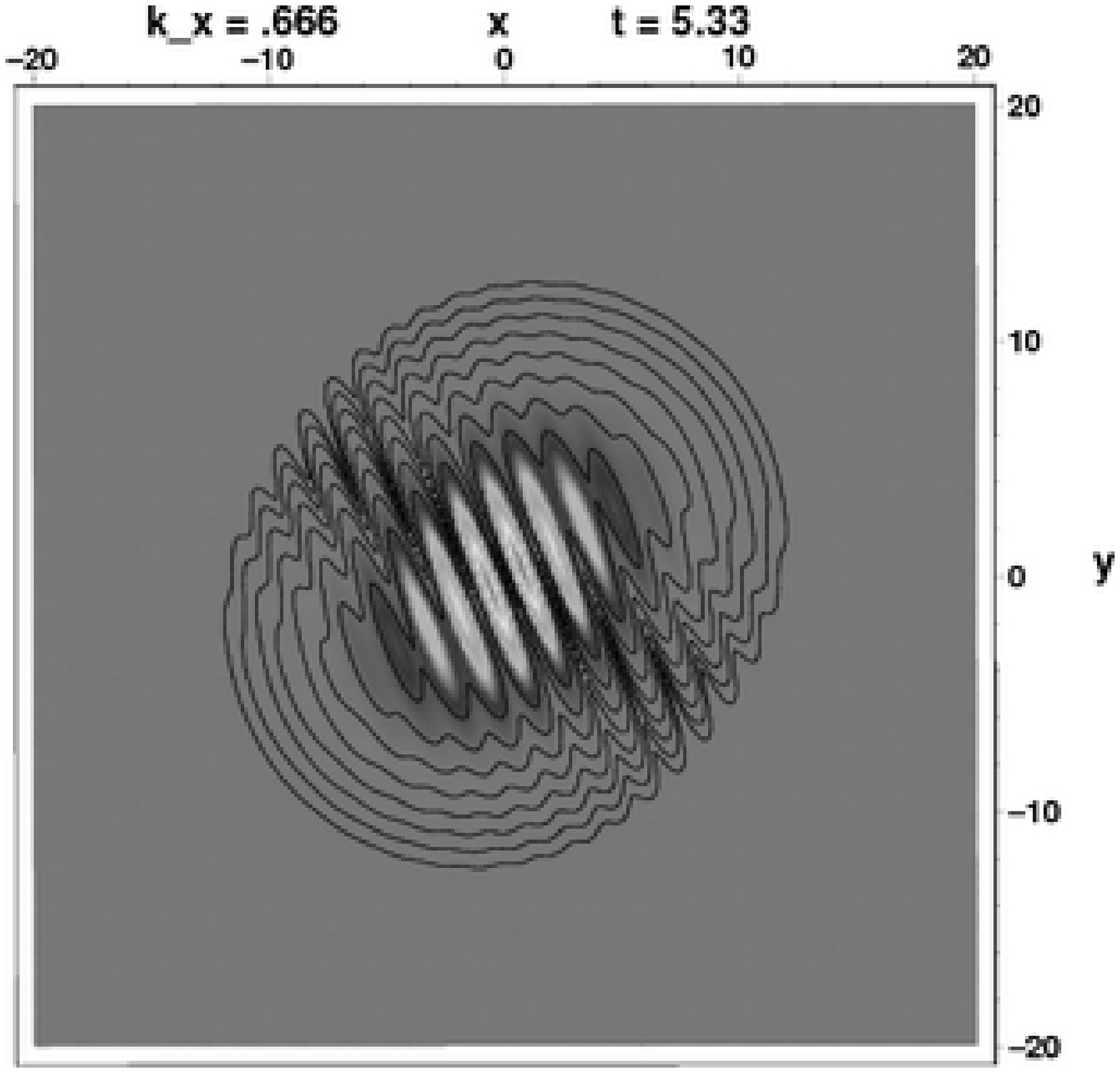}
\includegraphics[width=6cm]{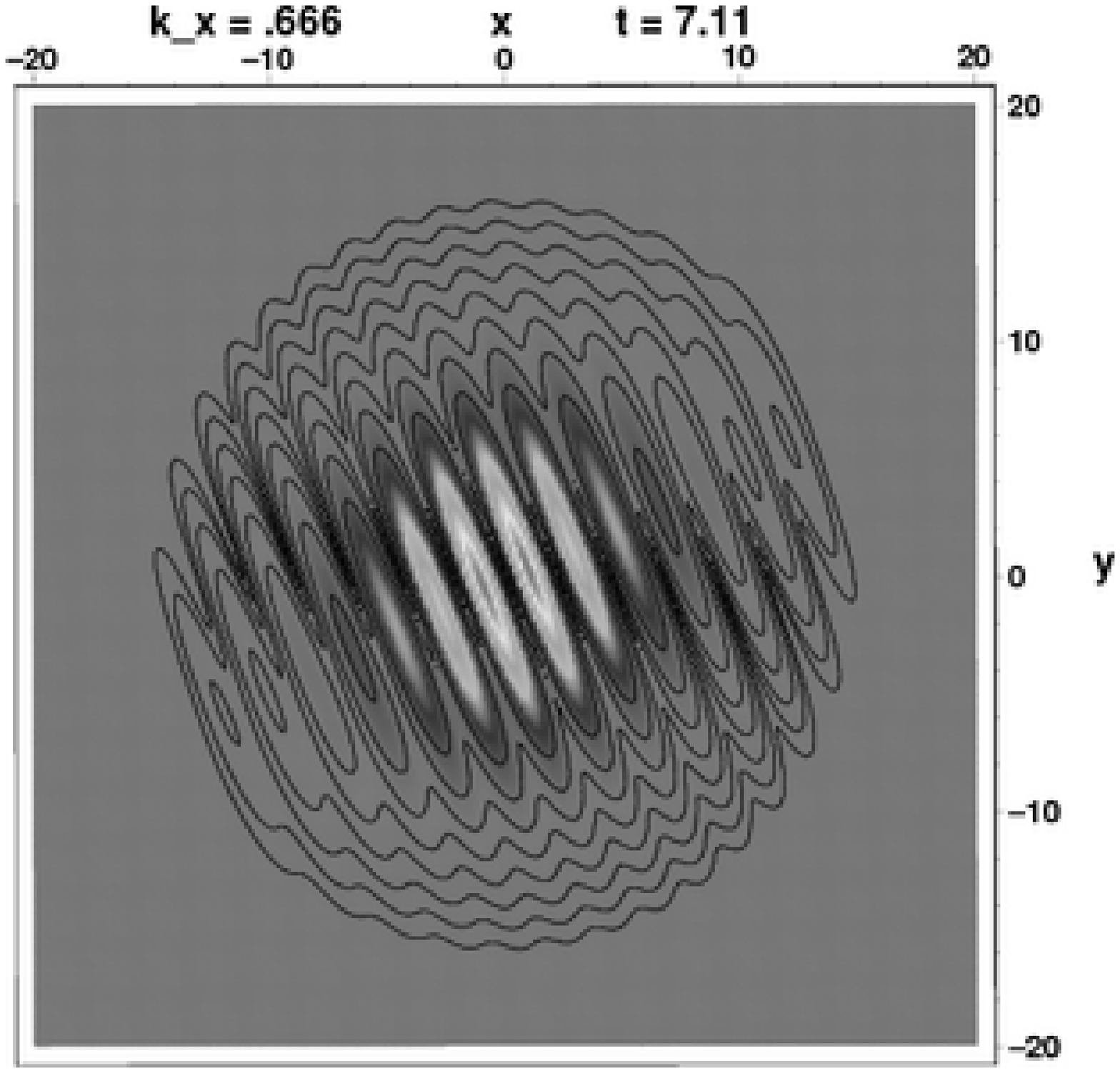}
\includegraphics[width=6cm]{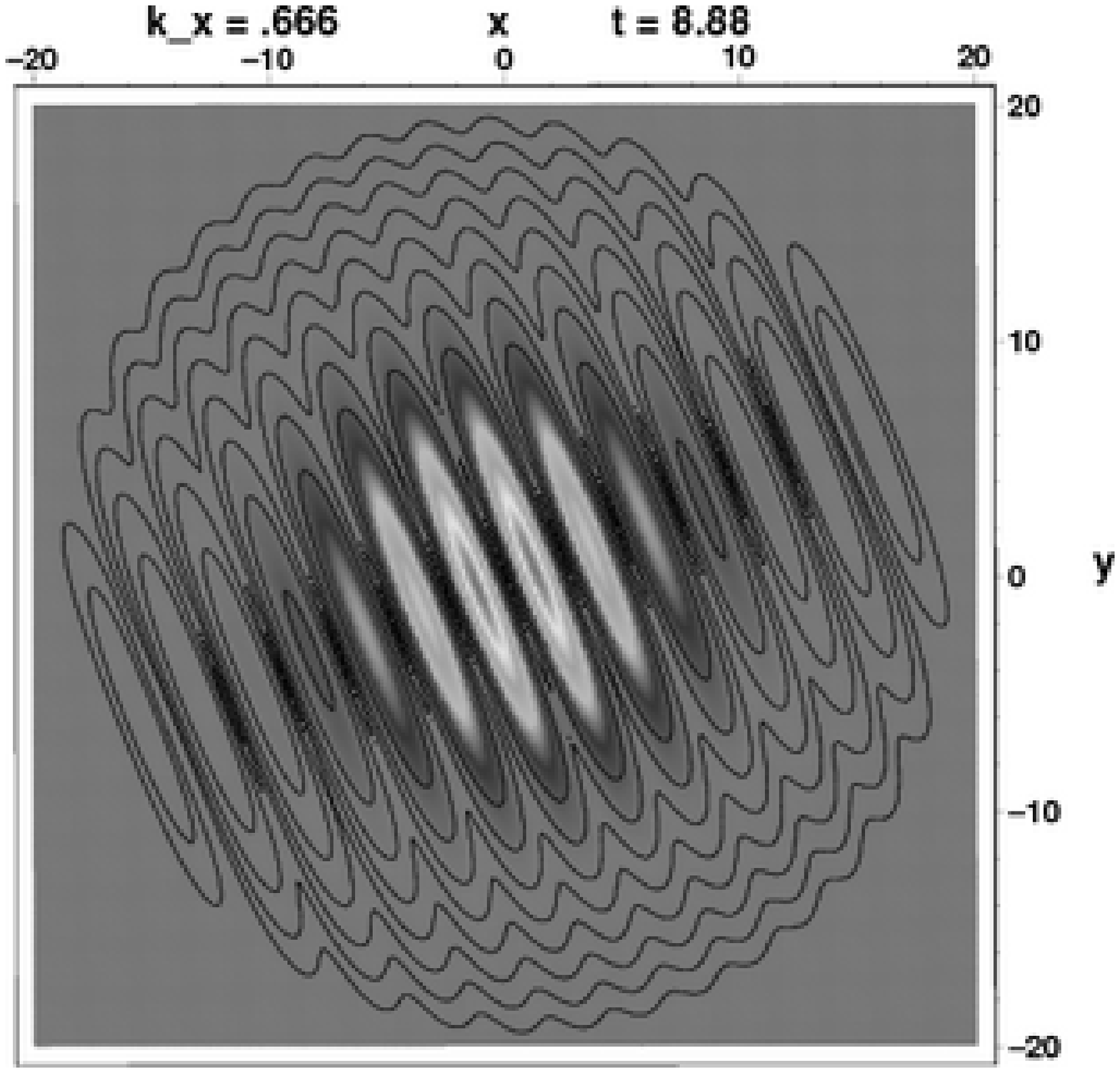}
\includegraphics[width=6cm]{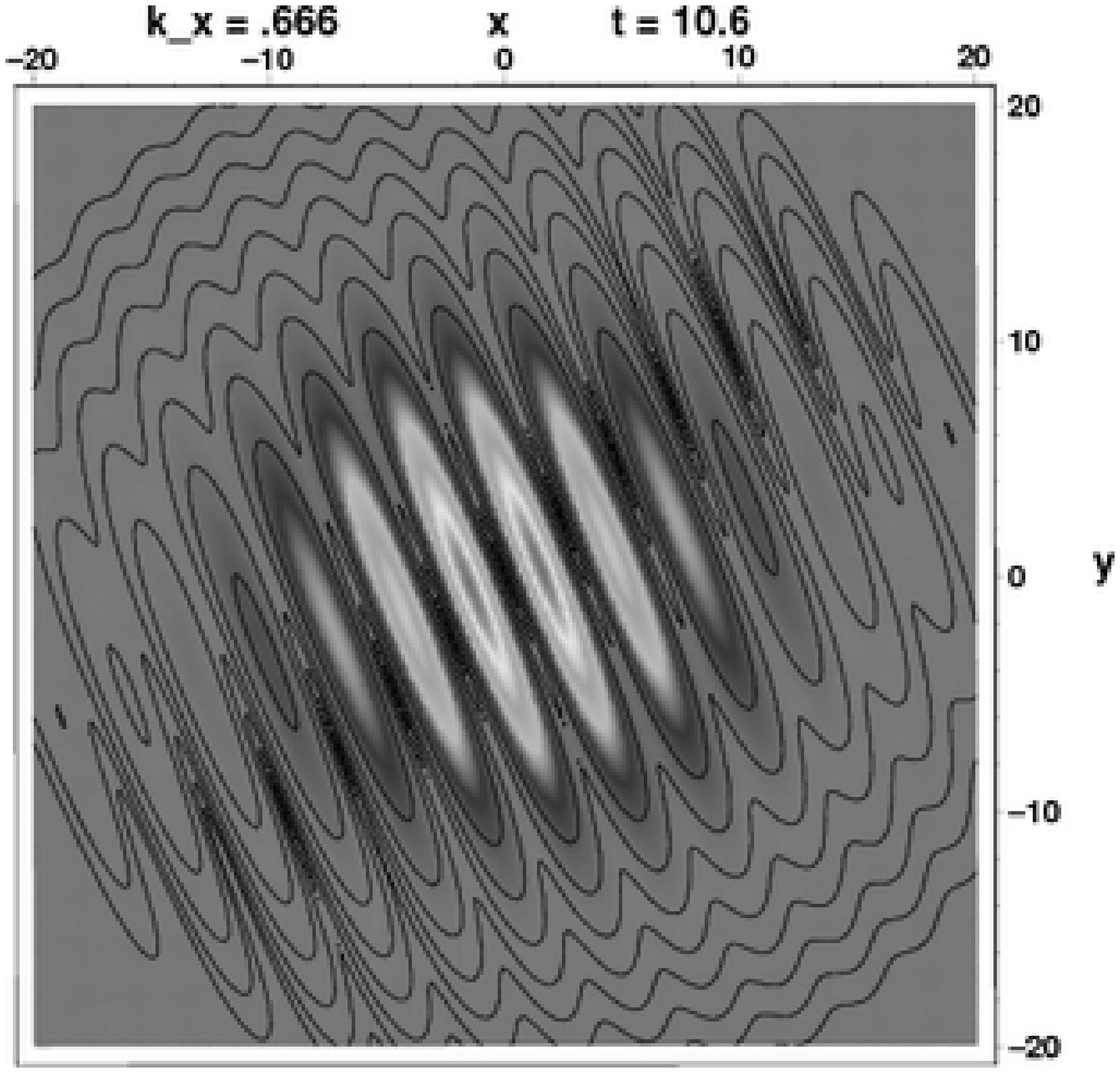}
}
   \caption{As in Fig.~(\ref{fig:classical}), but in the quantum
    regime ($k_{x}=2/3$)}
    \label{fig:quantum}
\end{figure*}
%%%%%%%%%%%%%%%%%%%%%%%%%%%%%%%%%%%%%%%%%%%%%%%%%%%%%%%%%%%%%%%%%%%%%%%%

\subsection{Interaction between two identical free fermions}

It is not difficult to solve exactly the time-dependent \Sch equation,
\begin{equation}
   i\hbar \,{\partial \over \partial t} \Psi(\vec x,t) =
       H \, \Psi (\vec x,t) \, ,
   \label{eq:schroedinger}
\end{equation}
for the Hamiltonian $H$ of two identical free fermions and illustrate
graphically the behaviour of the most physical object representing a
particle, the amplitude of the wave-function core.  We illustrate
below what happens to two spinless fermions in 3D space without
explicit interaction potential.  As initial condition we adopt
Gaussian wave-packets because they allow to calculate all the results
in closed form.

As derived in standard quantum mechanics, the full solution of the
single free particle \Sch equation in 3D Cartesian space reads (with
units normalised conveniently for our cosmological neutrino context to
$[l]=1\,{\rm cm}$, $[m]=0.1\,{\rm eV}$, and $[t]=1.7\cdot
10^{-7}\,{\rm s}$, which allows to set $\hbar/m=1$):
\begin{equation}
   \Psi (\vec x,t) = {1 \over (2\pi it)^{3/2}}
   \int\! \exp \left( i {(\vec x - \vec y)^2 \over 2 t} \right)
          \Psi(\vec y,0)\, d^3 y \, .
          \label{eq:sol}
\end{equation}
Thus finding how an arbitrary initial wave-function $\Psi(\vec x,0)$
evolves is just a matter of calculating a triple convolution integral
at any wished time $t$.  For particular initial $\Psi(\vec x,0)$ such
as Gaussian packets the integral can be completely expressed in closed
form, which we manipulate below with Maple.

When describing two independent distinguishable particles, the full
wave-function takes the form
\begin{equation}
   \Psi (\vec x_1, \vec x_2,t) =
        \Psi_1 (\vec x_1,t)\, \Psi_2 (\vec x_2,t) \, ,
\end{equation}
for which the time evolution is just the product of the single
particle solution Eq.(\ref{eq:sol}).

But this is only a very particular form of all the wave-functions
$\Psi (\vec x_1, \vec x_2,t)$ that the \Sch equation for two particles
can describe.  Functions that cannot be separated as a product of
single particle wave-function describe entangled states.  In
particular two identical particles must follow either a symmetric or
anti-symmetric wave-function upon the exchange of the particles.
These read
\begin{eqnarray}
    \nonumber
  \Psi_\pm (\vec x_1, \vec x_2,t) &=&  \\
  && \hspace{-8mm} 
   \frac{1}{\sqrt{2}}
      \big[\Psi_1 (\vec x_1,t)\, \Psi_2 (\vec x_2,t) \pm 
           \Psi_1 (\vec x_2,t)\, \Psi_2 (\vec x_1,t) 
      \big]  
      \, ,
\end{eqnarray}
where the $+$ sign applies to symmetric functions and the $-$ sign to
antisymmetric functions. The time evolution of such an entangled pair
of wave-functions can also be calculated exactly in closed form by
forming the product and sum of single particle solutions.  The
explicit formula is too long to show here, but can be easily
manipulated with Maple or similar software tools.

Although the wave-function of two particles is 6-dimensional and
complex, when the particle centres of mass and velocity are set at the
origin we can focus the attention on the function $\Psi_-(\vec x,
-\vec x,t)$ since any measurement of a particle at location $\vec x$
will constrain the position of the other particle at position $-\vec
x$, since we assumed to know the centre of mass.  $|\Psi_-(\vec x,
-\vec x,t)|^2$ describes the probability of finding one particle at
position $\vec x$ and the other one at $-\vec x$.

Below we show graphically the normalised value $|\Psi_-(\vec x, -\vec
x,t)|^2$ in the plane $z=0$ for the initial condition in the centre of
mass and velocity frame of two fermions represented initially by a
Gaussian wave-packet of unit half-width:
\begin{equation}
\Psi_1(\vec x, 0) = \exp(i \vec k\cdot \vec x) \,
\exp\left[-(\vec x-\vec x_0)^2/2\right] \ .
\end{equation}
The initial position of particle 1 is $\vec x_{1}(0)=\left\lbrace
  -5,-2,0\right\rbrace $, and the initial momentum is $\vec k_{1} =
\left\lbrace k_x,0,0 \right\rbrace $.  The initial position of
particle 2 is symmetric about the origin in order to put the centre of
mass and velocity at the origin.  The momentum is therefore all in the
component $k_x$, which is not directed toward the centre of mass in
order to illustrate a generic interaction with non-zero impact
parameter.

Here we show three different representative values of the momentum
$k_x$ illustrating the classical ($k_x=6$), semi-classical ($k_x=2$),
and quantum regimes ($k_x=2/3$).  This should suffice  for our purpose
of illustrating the exchange interaction of free identical particles.

In Fig.~\ref{fig:classical} we see the two wave-packets individually
spreading slowly due to their high momentum $k_x=6$.  Despite that, for
a while the packets strongly overlap, subsequently they recover their
localised wave-functions.  In such a case the individual packets
behave as expected for classical particles: they move in straight line
and preserve their identity.  Although Pauli's principle applies, it
has no practical consequence asymptotically.  These two particles can
be approximated as distinct again at later times.

In Fig.~\ref{fig:semi-classical} we see that at lower momentum,
$k_x=2$, two wave-packets interfere much more strongly.  The
antisymmetric wave-function always vanishes at the mass centre.  The
size of this exclusion region around the origin grows at lower
velocities.  The interference pattern presents elongated wave crests
which are inclined with respect to the original particle momentum
directions.  A plane wave seems to be building-up in a direction
slanted with respect to the particles directions.  Not only the
positions are strongly mixed in a single structure, but the momentum
becomes also mixed.  After a while the individual wave-packets
representing the classical localised particles succeed to partly
reform, but a remaining coma of entangled states remains around the
origin.

At still lower relative momentum, $k_x=2/3$, quantum effects become
dominant.  In Fig.~\ref{fig:quantum} the two particles initially
distinct rapidly loose their identity and form a single
anti-symmetrical wave-packet which spreads around the origin with waves
again reflecting an intermediate direction of propagation with respect
to the initial particle momentum.  Clearly in this case the
probability of finding a neutrino somewhere else than along the
initial straight trajectories is strongly enhanced by the quantum
interferences.  The classical approximation of straight trajectories
is no longer valid, and the initially distinct particles are entangled
forever.  In the quantum regime a clear dispersive behaviour occurs
with marked asymmetric past--future asymptotic states.

The same experiments have been repeated for bosons.  The only
notable difference is that the wave-functions are always positive
at the origin, and even take there maximum values in the quantum case:
this corresponds to the Bose-Einstein condensation. In contrast, the
fermionic wave-functions always vanish exactly at the origin, which
means that a volume of space around the origin is hardly available to
the particles. Fermi pressure is a consequence of confining 
the particle energy in a reduced volume.

\subsection{Interaction between $N$ identical free fermions}

The problem of describing statistically $N$ identical bosons or
fermions with or without interactions has been solved very early in
the history of quantum mechanics \citep{uhlenbeck32}, but is much more
difficult to illustrate graphically than the 2 particle case.  What is
clear is that in the limit of vanishing particle interaction, the
bosonic or fermionic statistical behaviour tends toward classical
Maxwell-Boltzmann statistic only at high temperature.  At sufficiently
low temperature the statistical behaviour tends toward the
Bose-Einstein and Fermi-Dirac statistics respectively.

%%%%%%%%%%%%%%%%%%%%%%%%%%%%%%%%%%%%%%%%%%%%%%%%%%%%%%%%%%%%%%%%%%%%%%%%
\begin{figure}
\resizebox{\hsize}{!}{\includegraphics{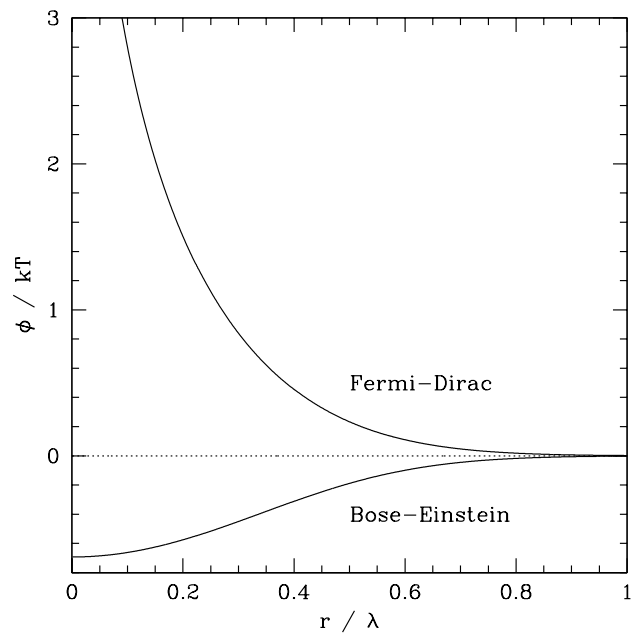}}
\caption{Effective particle-particle interaction potential $\phi$
  for a perfect gas of fermions (top curve, repulsive potential), or
  bosons (bottom curve, attractive potential) in first order of
  quantum correction, as a function of distance $r$ in unit of the de
  Broglie thermal wavelength $\lambda_{\rm dB}$.  }
\label{fig:pot}
\end{figure}
%%%%%%%%%%%%%%%%%%%%%%%%%%%%%%%%%%%%%%%%%%%%%%%%%%%%%%%%%%%%%%%%%%%%%%%%

An interesting aspect developed for example by
\citet[chap. 10]{huang64}, 
is that the first quantum correction to a classical perfect
gas, which is appropriate for our cosmological neutrino context,
caused purely by the bosonic or fermionic nature of the particles is
mathematically equivalent to an particle-particle interaction
potential, with the peculiarity to depend also on temperature $T$,
\begin{equation}
\phi(\vec x_i - \vec x_j) =
   -kT \log \left[ 1 \pm \exp\left( - 2\pi {|\vec x_i - \vec x_j|^2
   \over \lambda_{\rm dB}^2} \right)\right] \ .
\end{equation}
Fig.~(\ref{fig:pot}) shows that the pseudo potential is repulsive for
fermions and attractive for bosons.  The potential becomes important
($\phi \sim kT$) at a distance slightly below the de Broglie thermal
wavelength $\lambda_{\rm dB} = h/\sqrt{2\pi m k T}$.

This is consistent with the rule that quantum effects become
important when thermalised particles,
are less \textit{distant} in the classical description
than  $\lambda_{\rm dB} $. The potential permits to derive an
effective interaction cross-section, and from there the time required
for particle to become entangled.  We will refer to this time as the 
coherence time-scale.  If the particles were classical we would call
this a collision time-scale, that would determine a relaxation time.
 
If we adopt a classical description of neutrinos moving in such a
potential, clearly a strong deflection of trajectory occurs whenever
$\phi/kT \sim 1$, which means also that $|\vec x_i - \vec x_j| \sim
\lambda_{\rm dB}$, so the interaction cross-section reads $\sigma \sim
\pi\lambda_{\rm dB}^2$.  Thus the first order quantum correction to a
classical description of semi-degenerate neutrinos introduces an
interaction potential producing ``entanglement'' with a 
characteristic time-scale given by
\begin{eqnarray}
\tau \ \approx \ {1 \over n_\nu \sigma v_\nu} 
     & \approx & { {m_{\nu}^{3/2} \sqrt{kT}} \over { n_\nu h^2}} \nonumber \\
     & \approx & 
    1.3\cdot10^{-8} (1+z)^{-2} \left(n_{\nu,0} \over 56 \right)^{-1}
                    \left(m_{\nu} \over 1{\rm eV} \right)^{-1} \, {\rm s}
\ ,
\end{eqnarray}
using the asymptotic relations for $T$, $n$, and $v$ given in Sect.~3.3. 
This yields a very short time-scale $\ll 10^{-6} \, \rm s$ for any
reasonable present day and past cosmological neutrinos.  Although this
derivation is not consistent since quantum particles do
not follow classical trajectories, at least $\tau$
characterises the entanglement time-scale between neighbour
neutrinos. 

\subsection{Decoherence time}
In the decoherence theory \citep{zeh70,zeh05,zurek02,zurek03},
classical physics emerges from the much larger sensitivity of the
entangled states in the density matrix $\Psi(x) \Psi^*(x^\prime)$,
$x\neq x^\prime$, to the interferences coming from the outer world
than the diagonal states $\Psi(x) \Psi^*(x)$.  Particles become
localized classical objects because the uncorrelated perturbations
from the outside world prevent the wave-function to spread.  The
decoherence time $\tau_D$ for a quantum system considered over a
distance $\Delta x$ to loose quantum correlations $\Psi(x)
\Psi^*(x^\prime)$, $\Delta x < |x-x^\prime|$, is related to the
relaxation time $\tau_R$ induced by the outer world, and to
$\lambda_{\rm dB}$:
\begin{equation}
  {\tau_D \over \tau_R} = 
  \left(  \lambda_{\rm dB} \over \Delta x \right)^2
  \ .
\end{equation}
Thus the decoherence time is inversely proportional to the square of
the considered region.  Obviously, to quantify decoherence we must
identify the fastest process that perturb or ``relax'' neutrinos.

Let us make a best effort to find the shortest possible $\tau_R$ that
cosmological neutrinos might be subject to.  If we discard weak
interaction processes (cf.~Eq.~(\ref{equ:WS})), we only have
gravitational deflection available.

Suppose we have a density $n$ of bound gravitational objects of mass
$M$.  The gravitational cross-section for neutrinos at velocity
$v_\nu$ is $\sigma = \pi G^2 M^2/v_\nu^4$, and the timescale for
deflecting significantly neutrinos by these objects is:
\begin{equation}
  {\tau_R} = {1 \over n \sigma v_\nu} 
           = {v_\nu^ 3 \over \pi G^2 n M^2 } \ .
\end{equation}
Thus at constant mass density $nM$ more massive objects deflect faster
than light objects.  Galaxies seem the most promising deflectors,
because stars are mostly bound in galaxies and amount to only a fraction
of their mass, while galaxy clusters contain a small fraction of the
galaxies.

Plugging in plausible numbers for galaxies ($n \approx 10^{-74}\,\rm
cm^{-3}$, $M \approx 10^{45}\,\rm g$), and $v_\nu \approx 10^8 \,\rm
cm\,s^{-1}$, we obtain $\tau_R \sim 10^{22}\, \rm s$.  Therefore for
$\tau_D$ to be less than the entanglement time $<10^{-6}\,\rm s$, we
need to consider regions larger than
\begin{equation}
  \Delta x =  \lambda_{\rm dB} \sqrt{\tau_R \over \tau_D} 
  \approx 10^{13} \, {\rm cm} \ ,
\end{equation}
in order to begin to use classical physics.  In such a region of AU
dimension we have $\sim 10^{40}$ neutrinos that behave as a fermionic
ensemble.  Since the relaxation time $\sim 10^{22}\, \rm s$ is much
larger than the neutrino age $\sim 10^{17}\, \rm s$, most neutrinos
that would be localized by hypothetical detectors would still be
entangled with neutrinos at cosmic distances apart. The ratio of these
times, $10^{-5}$, gives the fraction of neutrinos that have been
perturbed by the gravitational attraction of a galaxy.  After $10^5$
entanglement times of $<10^{-6}\, \rm s$, each neutrino has a
large probability to be also entangled with perturbed neutrinos, thus
propagating the coherence loss.

%_______________________________________________________________________

\section{Discussion}

\subsection{The status of quantum physics}

The quantum physics rules are so strange that it took a long time for
the physicist community to make a standard view about its principal
aspects.  Today several points are not completely resolved, such
as the meaning of the measurement process, or the r\^ole of time in a
relativistic quantum theory of several particles \citep{penrose05}.
Quantum gravity is still in the work.

However in the last decade many new experiments have confirmed the
strange rules of quantum mechanics. The Einstein-Podolsky-Rosen (EPR)
paradox has been confirmed, two entangled particles can fly apart over
long distances and preserve their mutual correlation alive even when
one
of the particle is subject to measurement.  
In recent experiments \citep{tittel98}
the photon entanglement has been preserved over distances of the order
of $10\,\rm km$ throughout a complicated network of telecommunication
optical fibres.

\subsection{Hanbury Brown and Twiss (HBT) effect and 
the quantum decoherence}

At the root of quantum optics, there is a fine example of exchange
symmetry interaction operating on otherwise \textit{interactionless}
particles: photons.  This effect was first investigated by Hanbury
Brown and Twiss \citep{hanbury54} with the now famous intensity
interferometer \citep{hanbury74}.  The effect could also be performed
at the laboratory scale \citep{morgan66} and was an important step in
the development of quantum optics.  In the Hanbury Brown and Twiss
experiment, independent incoherent photons emitted all over the
surface of a star ($D \sim 10^9 \, \rm m$) at distinct times are
collected on Earth by two distinct telescopes separated by macroscopic
distances ($d \sim 10^2\, \rm m$). The collected distinct photons
arriving in the two detectors within a photon coherence time are
verified to be \textit{correlated in time}. Despite being particles
without explicit coupling term and initially incoherent, photons
sufficiently close in phase space (they are very close in momentum
space) succeed to bunch, i.e., to exchange momentum.  This is an
example of a partial Bose-Einstein condensation.

The fermion effect corresponding to the HBT experiment has
been confirmed recently for electrons in laboratories
\citep{henny99, kiesel02}.  Such experiments are complicated by the
fact that the repulsive electro-magnetic interaction term of electrons
must be cancelled by either an atom lattice, or a sufficiently large
distance between electrons.  In these experiments, electrons
overlapping their wave-functions are found to be anti-correlated, which
is at the root of the degeneracy pressure in fermionic gases.  Better
suited particles are neutrons, for which the HBT effect has also been
verified recently \citep{iannuzzi05}.

The important point of these verified phenomena is to show that
identical particles interact also through their exchange symmetry, and
that the speed of this kind of interaction, not explicit in the
Hamiltonian, is in practice determined by the speed at which the
core of the wave-functions overlap significantly.

If we take the strong overlap of the particle wave-functions as the
principal requirement for particles to be said \textit{interacting},
and the weak overlap of many uncorrelated wave-functions as the regime
during which the decoherence due to the rest of the world acts, we can
understand why our common experience of the world is not a fully entangled
structure \citep{penrose05}, but the classical world that we know,
where spatially distinct regions and classical devices 
can be assumed to exist. 
Correlations between particles are constantly lost by decoherence and
built by wave-function overlaps.  The macroscopic result for
multiparticle systems is equivalent to a relaxation toward a thermal
state \citep{gemmer04}.

Therefore if we apply these considerations to cosmological neutrinos,
these must overlap their wave-function very frequently ($\ll
10^{-6}\,\rm s$), and each ``interaction'' produces a partial entanglement
of the wave-functions. Quickly neutrinos get entangled to a growing
number of other neutrinos which together form a spreading entangled
wave-function.  Any perturbation or ``observation'' of a given
neutrino entangled with others concerns then all the other neutrinos.
Everything happens as if the detected neutrinos had exchanged momentum 
with the others, like classical particles in usual interactions do.

%_______________________________________________________________________

\section{Conclusions}

In view of the solidity of quantum mechanics and the likely sub-eV
mass of neutrinos, we arrive to the conclusion that quantum effects
play an dominant r\^ole for cosmological neutrinos.  Despite being
almost interactionless, these neutrinos form a non-localisable
ensemble of entangled particles subject to Pauli's principle.  The
weakness of their interactions with the outer world is precisely the
key factor that prevents them to decohere faster than they become
entangled.  The only way to justify a classical description of
neutrinos following the collisionless Boltzmann's equation would be to
find a perturbative process that would decohere all of them in much
less than $10^{-6}\,$s.  But then this perturbative process should 
presumably be included in the neutrino dynamics.

With the most accurate description of particles, the wave-function, we
have illustrated explicitly how a pair of initially localised
neutrinos becomes entangled fast due only to their antisymmetry;
quickly the localised wave-packets become a single fully non-localised
object.  In practice this means that more neutrinos act rapidly as an
ensemble that must be treated with the standard quantum statistics
tools.  Since, except for antisymmetry, no interaction potential is
significant, the appropriate description at cosmological scales is
simple, the one of a semi-degenerate perfect fermion gas.

For structure formation, the Fermi pressure contribution has little
importance as long as the Universe is homogeneous, since only pressure
gradients play a dynamical r\^ole.  At redshifts $z \gg 10^5$ the
Fermi over-pressure amounts to 5\% of the kinetic pressure.  But at
redshifts $z \ll 10^4$ in the non-relativistic regime, during the
non-linear structure formation epoch, the over-pressure increases to
68\%.  Then a Fermi gas of massive neutrinos is substantially
different from a collisionless gas.  Not only the velocity dispersion
remains then isotropic in a tightly coupled gas, but since collisional
fluids are able to shock due to the non-linear convective term in the
fluid equations, the evolution of the combination of neutrino gas,
dark matter, and baryons in the today Universe may turn out to be
complex.  Shocks typically increase entropy, degeneracy, and therefore
pressure, as seen in Sect.~3.  It would be interesting to determine
how far neutrinos could violate the phase space density conservation
by dissipative processes, especially when mixed with baryons.

The extension of these results to other dark matter particle
candidates resulting from the Big Bang should be straightforward.
Relativistic fermions, degenerate or not, can not contribute
significantly to matter clustering.  Usual heavy cold dark matter
fermions, such as neutralinos, are assumed much less dense and much
heavier than neutrinos, in which case phase space degeneracy may
remain small and decoherence fast, allowing the use of the
collisionless fluid approximation.

\begin{acknowledgements}
We thank particularly Nicolas Gisin, Thanu Padmanabhan, and Jack
Steinberger for sharing their point of views and insight of
physicists.  We have appreciated private discussions with Joel Primack
and Ben Moore about the cosmologist view. We enjoyed discussions with
Silvia Pascoli and Mario Campanelli at CERN about recent experimental
results on neutrino physics.  We thank our colleagues Andr\'e Maeder
and Denis Puy for lively discussions too.  This work has been
supported by the Swiss National Science Foundation.
\end{acknowledgements}

%_______________________________________________________________________

\end{document}